\documentclass[preprint,12pt,3p]{elsarticle}

\usepackage[colorlinks=true, citecolor=magenta, linkcolor=blue, urlcolor  = blue]{hyperref}  
\usepackage{amsmath,amssymb,amsfonts,amsthm}
\usepackage{longtable}
\usepackage{graphicx}
\usepackage{booktabs}
\usepackage{mathtools}
\usepackage{amsmath}                
\usepackage{amsthm}                
\usepackage{amssymb}

\theoremstyle{remark}

\theoremstyle{definition}
\newtheorem{defn}{Definition}[subsection]

\journal{Renewable and Sustainable Energy Reviews}

\begin{document}

\begin{frontmatter}

\title{An Enterprise Control Assessment Case Study of the Energy-Water Nexus for the ISO New England System}

\author[label2]{Steffi O. Muhanji\corref{cor1}}
\address[label2]{Thayer School of Engineering, Dartmouth College, Hanover, NH 03755}

\cortext[cor1]{I am corresponding author}

\ead{Steffi.O.Muhanji.TH@dartmouth.edu}

\author[label4]{Clayton Barrows}
\address[label4]{National Renewable Energy Lab (NREL), Golden CO, (United States)}
\ead{clayton.barrows@nrel.gov}

\author[label4]{Jordan Macknick}
\ead{jordan.macknick@nrel.gov}

\author[label2]{Amro M. Farid}
\ead{Amro.M.Farid@dartmouth.edu}

\begin{abstract}
The electric power generation mix of ISO New England (ISO-NE) is fundamentally changing. Nuclear, coal, and oil generation facilities are retiring while natural gas, solar, and wind generation are being adopted to replace them. Variable renewable energy (VREs) such as solar and wind present multiple operational challenges that require new and innovative changes to how the electricity grid is managed and controlled. Within the context of a New England case study, this paper looks at ways in which the water supply systems (water and wastewater treatment), and water dependent electricity generating resources (hydro, and thermal power plants) can be operated flexibly to help balance energy in an evolving grid. The study's methodology employs the novel but now well published Electric Power Enterprise Control System (EPECS) simulator to study the electric power systems operation, and the System-Level Generic Model (SGEM) to study the associated water consumption and withdrawals. This work studies six potential 2040 scenarios for the energy-water nexus of the ISO-NE system.  It presents a holistic analysis that quantifies power system imbalances, normal operating reserves, energy market production costs, and water withdrawal and consumption figures.  For scenarios with a high penetration of VREs, the study shows great potential of water resources to enhance grid flexibility through the provision of load-following, ramping, and regulation reserves by water resources.  The work also provides significant insights on how to jointly control the water and energy supply systems to aid in their synergistic integration. 
\end{abstract}

\end{frontmatter}

\section{Introduction}
\begin{figure}[!ht]
\centering
\includegraphics[width=6.2in] {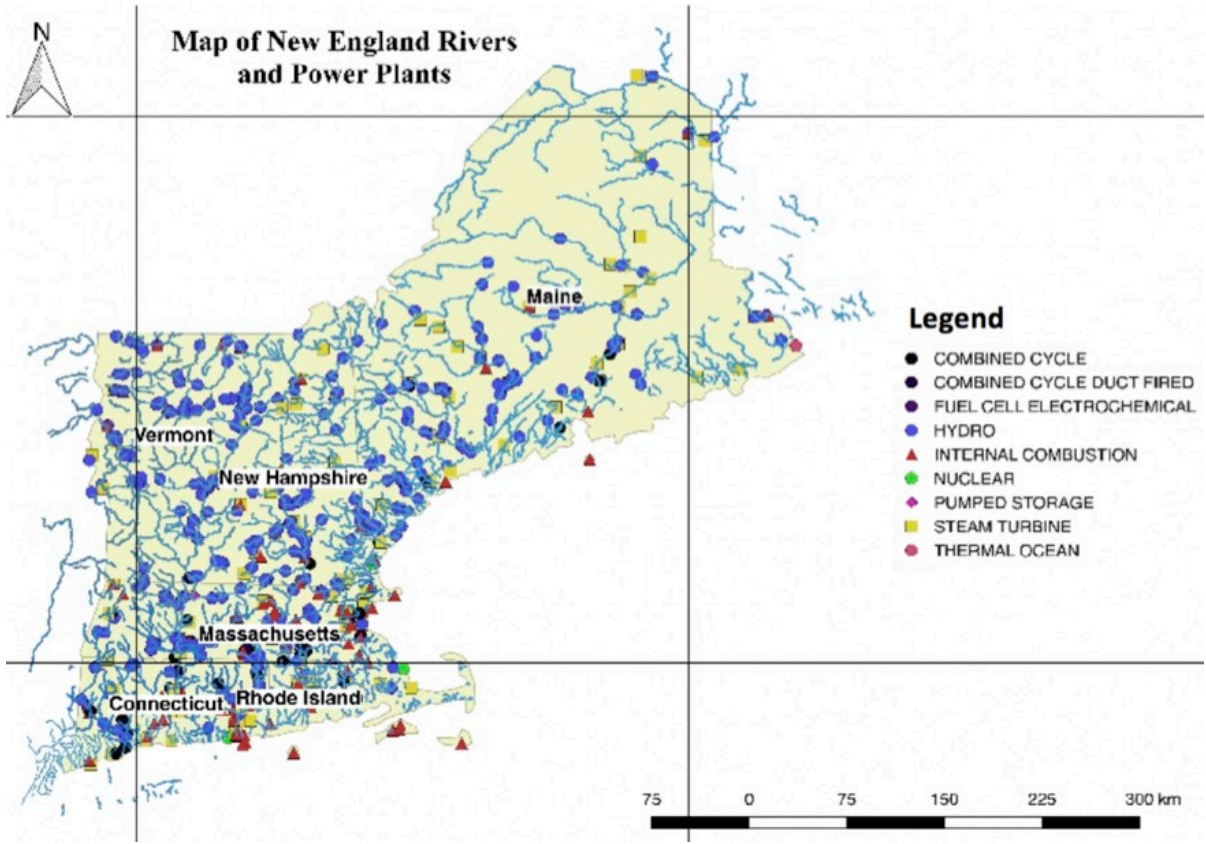}
\caption{A map of New England's electric power generation units and rivers.} 
\label{fig:waterelecInfra}
\end{figure}

The bulk electric power system of New England is fundamentally changing to include more solar and wind generation resources. This evolving resource mix has triggered changes to how the power grid is managed and controlled. The bulk of these changes have been in capacity and transmission expansion. However, with the growing uncertainty and variability introduced by variable renewable energy, there is an even greater need for increased amounts of operational flexibility \cite{Farid:2016:00,EPRI:2016:00}. Water plays a fundamental role in the ISO New England system. Conventional and run-of-river hydro make up over 9\% of the overall generation in the 6 New England states\cite{ISO-NE:2017:02}. An additional 1\% of generation comes from the two main pumped-water storage facilities, Bearswamp and Northfield\cite{ISO-NE:2017:02}. In the meantime, over 83\% of the current ISO-NE generation fleet comes from thermal generation facilities which withdraw and consume large quantities of water for cooling purposes\cite{ISO-NE:2017:02}. In spite of the changing resource mix, recent studies predict that thermal generation facilities will still account for a significant percentage of future generation facilities in 2040\cite{Sanders:2014:00,Armstrong:2018:00}. Fig.~\ref{fig:waterelecInfra} illustrates the extent of the coupling between the water and electricity generation resources in New England. From Fig.~\ref{fig:waterelecInfra}, it is clear that most generating facilities are located near a water source and rely on adequate water supply to perform their function. These factors not only indicate significant coupling between the water and electricity supply systems but they also emphasize the need for more coordination between the two systems. Specifically, the potential synergies between the two systems cannot be ignored especially as the electricity grid undergoes its sustainable energy transition. 

\begin{figure*}[!ht]
\centering
\includegraphics[width=6.2in]{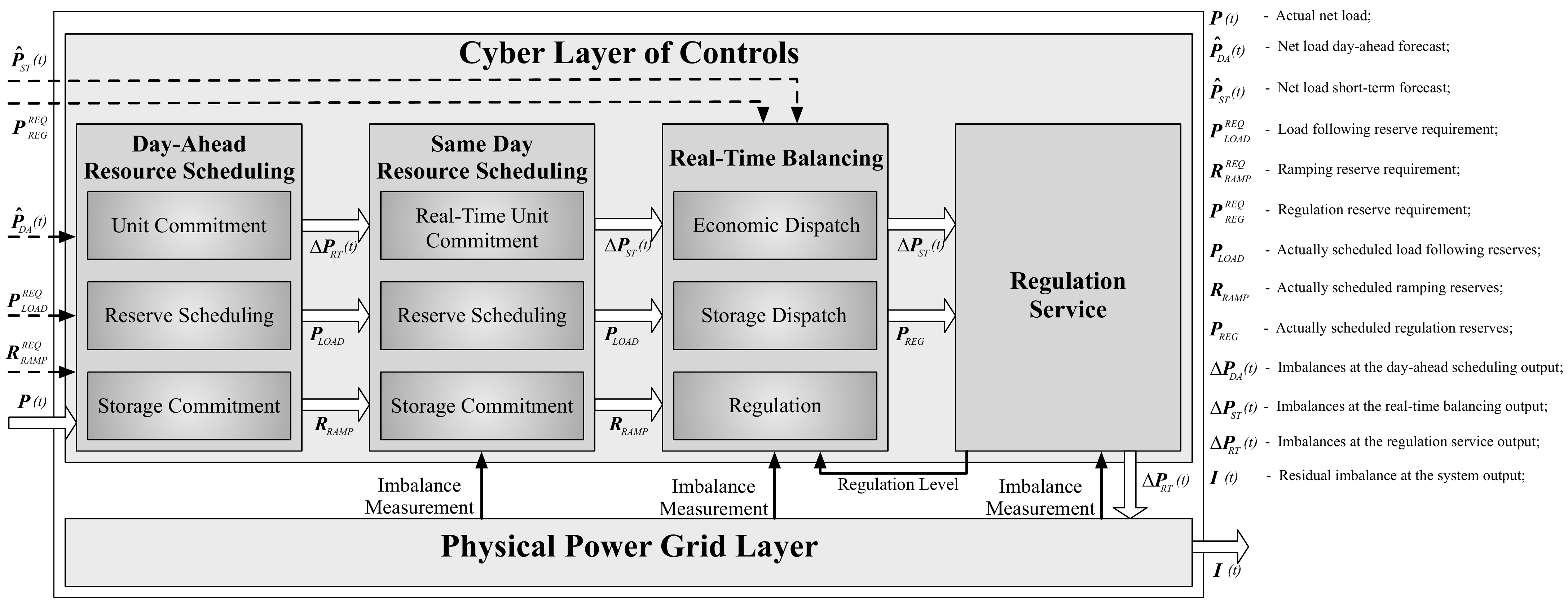}
\caption{Architecture of the Electric Power Enterprise Control System (EPECS) simulator customized for ISO New England operations \cite{Muzhikyan:2019:SPG-JR04}}
\label{fig:epecs}
\end{figure*}
 Concern about water security is growing especially with climate change affecting hydrology patterns and the decline of freshwater resources\cite{Rogers:2013:00,Kanyerere:2018:00,Averyt:2013:00}. At the same time, significant attention has gone into the integration of variable renewable energy into the electricity grid as a means of decarbonizing the electricity supply system. As discussed in the prequel to this paper\cite{Muhanji:2019:SPG-JR05}, the challenges of renewable energy integration and energy-water-nexus are very much related. In addition to presenting low $CO_2$ emissions, VREs have very low life-cycle water intensities\cite{Al-Nory:2014:00}. On the other hand, water is easily stored and therefore, has the potential to serve as a flexible energy-water resource on both the supply-side as well as the demand-side\cite{Lubega:2016:00}. As a result, the methodologies of energy-water-nexus and renewable energy integration studies must converge in order to realize potential synergies.

With the growing penetration of wind and solar generation in the ISO-NE system, grid operators would benefit immensely from an increased number of flexible resources that can be used to balance the grid in the real-time.  Similarly, water system operators could offer ancillary services, improve their profits and also achieve a more robust operation of their systems. Despite these benefits, renewable energy integration and EWN studies have not yet converged to realize benefits. While some energy-water nexus studies have quantified the withdrawals by thermal power plants, these studies have largely been conducted in isolation of actual operation of the electricity generation industry\cite{Rogers:2013:00,Averyt:2013:00,Meldrum:2013:00,Macknick:2012:00,Averyt:2011:00,Macknick:2012:01}. Thus, the full impact on either infrastructure is not assessed. Other EWN works have focused solely on optimizing the operations of water systems such as in the optimal operation of water pumps and optimal pump scheduling\cite{Bagloee:2018:00,Bagirov:2013:00,Ulanicki:2007:00,Lopez-Ibanez:2008:00,Ghelichi:2018:00} in order to provide demand response and other ancillary services while maximizing returns for water system operations\cite{Diaz:2017:00,Takahashi:2017:00,Menke:2017:00,Menke:2016:01}. Finally, a small subset have presented mostly single-layer approaches to co-optimize the water and electricity networks. Examples of such works include the optimal network flow in \cite{Santhosh:2013:EWN-C16}, the economic dispatch in \cite{Santhosh:2012:EWN-C09}, and the unit commitment problem in \cite{Hickman:2017:EWN-J32} for a combined water, power, and co-production facilities. Despite the large body of work and research on the energy water nexus, there is still a lack of a generic, case and geography-independent methodologies that encompass all flows within, and between the water and energy systems.

On the other hand, renewable energy integration studies have often been case and geography specific and have mostly utilized unit-commitment-economic-dispatch (UCED) models of power system control to study the operation of electricity markets with large penetrations of VREs\cite{Ela:2009:00,Brouwer:2014:00,Holttinen:2012:01,Holttinen:2013:00,Muzhikyan:2019:SPG-JR04}\cite{GE-Energy:2010:01,Shlatz:2011:00,Piwko:2005:00}. A significant percentage of these studies have taken statistical approaches to determine the impact of wind and solar forecast errors on dispatch decisions. A majority of renewable integration studies have recognized the vital role of reserves in the balancing performance of systems with high VRE penetration and have thus focused on the acquisition of normal operating reserves such as load-following, regulation, and ramping reserves\cite{Ela:2009:00,Brouwer:2014:00,Holttinen:2012:01,Holttinen:2013:00,Muzhikyan:2019:SPG-JR04}. 

However, a recent review of renewable integration studies shows major methodological limitations\cite{Farid:2014:SPG-J26}. Firstly, while some studies focus on reserve acquisition, the required quantity of reserves is usually based on the experiences of grid operators which no longer applies to systems with high penetrations of VREs\cite{Holttinen:2008:01,Robitaille:2012:00}. Secondly, most studies only consider either the net load variability or the forecast error in determining the amount of reserves despite evidence that shows that both of these variables contribute towards normal operating reserve requirements\cite{Ummels:2007:00,Holttinen:2008:01}. Lastly, although studies have shown that VREs possess dynamics that span multiple timescales of power system operation\cite{Curtright:2008:01,Apt:2007:00,Milano:2010:17}, most renewable energy integration studies have largely neglected the effect of timescales on the various types of operating reserve quantities\cite{Farid:2014:SPG-J26}.  Farid et al. \cite{Farid:2014:SPG-J26} proposed a holistic approach based on enterprise control to study the full impact of VREs on power system balancing performance and reserve requirements while considering the multi-timescale dynamics of VREs. \emph{Enterprise control} is an integrated and holistic approach that allows operators to study and improve the technical performance of the grid while realizing cost savings\cite{Farid:2014:SPG-J26}. An application of enterprise control in the form of the Electric Power Enterprise Control System (EPECS) simulator has been proposed in literature\cite{Muzhikyan:2017:SPG-J36,Muzhikyan:2015:SPG-J22,Muzhikyan:2015:SPG-J16,Muzhikyan:2015:SPG-J15,Muzhikyan:2014:SPG-C43,Muzhikyan:2014:SPG-AP07} and tested on various case studies including the ISO New England system\cite{Muzhikyan:2019:SPG-JR04}. In \cite{Muzhikyan:2019:SPG-JR04}, the EPECS simulator is used to study the performance of the ISO-NE system on 12 scenarios with varying penetrations of VREs. This study highlights the key role of curtailment and normal operating reserves on the balancing performance of the ISO-NE system. This paper extends the work in \cite{Muzhikyan:2019:SPG-JR04} and \cite{Muhanji:2019:SPG-JR05} to quantify the flexibility afforded the ISO-NE system through flexible operation of water resources.
\begin{figure}[!ht]
\centering
\includegraphics[scale=0.5]{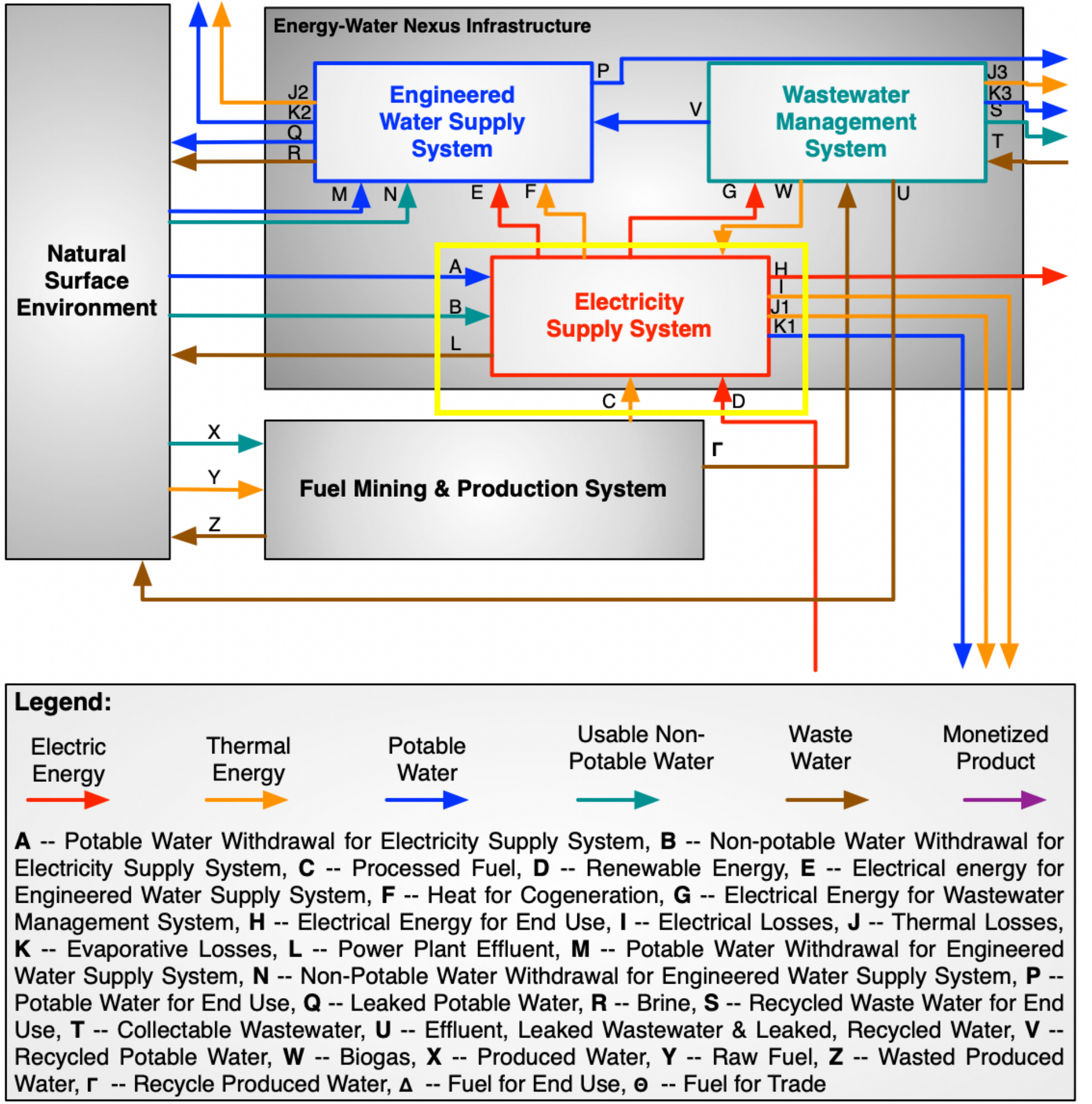}
\caption{A diagram of the physical flows between the physical infrastructures (water supply system, wastewater management system, and electricity supply system) and the natural surface environment. \cite{Abdulla:2015:EWN-C53}}
\label{fig:ewnflowchart}
\end{figure}
\subsection{Contribution}
This paper applies the methodology presented in the prequel \cite{Muhanji:2019:SPG-JR05} and \cite{Muzhikyan:2019:SPG-JR04}  to study the techno-economic performance of the energy-water-nexus for the ISO-NE system focusing on six predefined scenarios in 2040. The study methodology takes the {\color{blue}{yellow}} rectangle of Fig.~\ref{fig:ewnflowchart} as its system boundary. Given this specific choice of system boundary, this study is able to quantify the mass and energy flows in and out of the defined system boundary regardless of the test case or geography. The paper also provides insight into some of the operational challenges presented by high penetrations of VREs and also quantifies the amounts of normal operating reserves for the ISO-NE system for each scenario. Given that the methodology presented in the prequel \cite{Muhanji:2019:SPG-JR05} is generic and modular, the EPECS simulator is modified slightly to reflect the ISO-NE methodology (fully presented in \cite{Muzhikyan:2019:SPG-JR04} and as shown  in Fig.~\ref{fig:epecs}). In this study, the following quantities are studied: 1) load-following, ramping, and regulation reserves, 2) the demand response potential of water units, 3) the fuel flows of thermal units and their carbon emissions, 4) water withdrawals and consumption by thermal power plants, and 5) the effect of flexible operation of water resources on the production cost operation of the New England electricity grid.

\subsection{Outline}
The rest of the paper is structured as follows: Section \ref{sec:Methodology} presents the methodology for the ISO New England Energy-Water Nexus study. Section \ref{sec:casestudy} gives a detailed description of the case study data. Section \ref{sec:results} presents the results of the study within the context of the key performance characteristics of the power grid. Finally, the paper is concluded in Section \ref{sec:conc}.
\section{Methodology}\label{sec:Methodology}
\begin{figure}[!t]
\centering
\includegraphics[scale=.65]{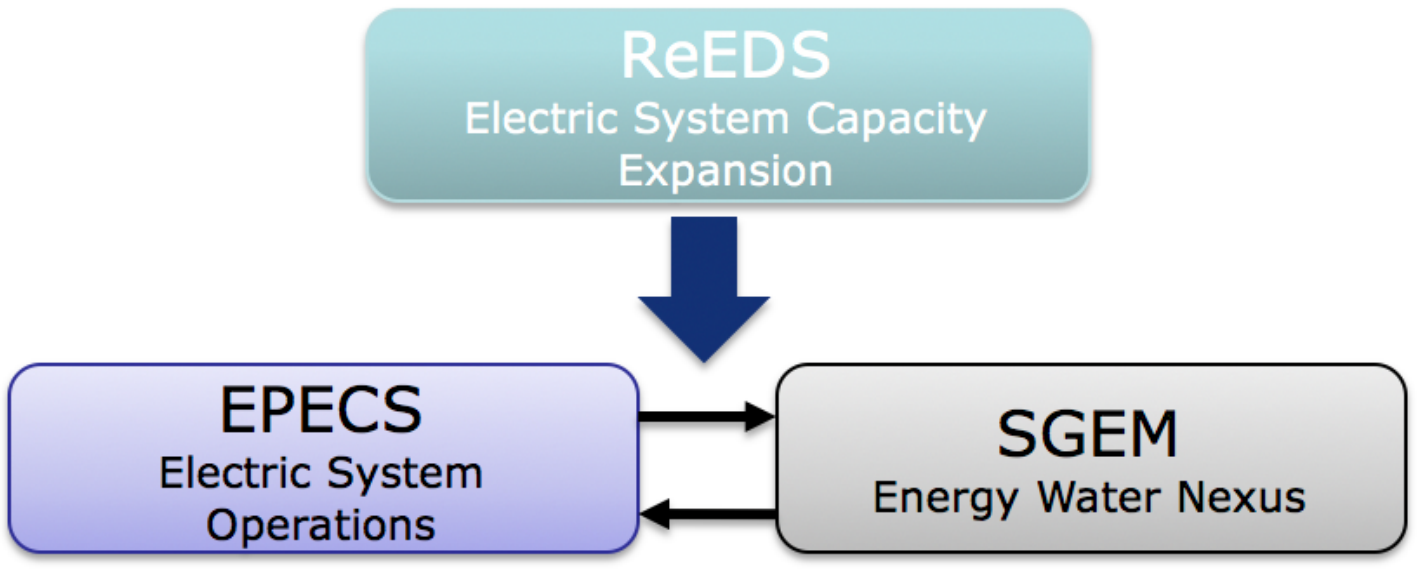}
\caption{Diagram of the Simulators Used in the ISO New England Energy-Water Nexus Study.}
\label{fig:methodologySimulators}
\end{figure}

As shown in Figure \ref{fig:methodologySimulators}, the methodology of the ISO New England Energy-Water Nexus study is best viewed in two parts:  planning and operations.  Section \ref{sec:planning} describes how the National Renewable Energy Laboratory's (NREL) Regional Energy Deployment System (ReEDS) was used to evolve the 2030 ISO New England electric power generation capacity mixes to six distinct 2040 capacity mix scenarios.  From there, the remainder of the section describes the Electric Power Enterprise Control System (EPECS) simulator as customized for ISO New England's operation\cite{Muzhikyan:2019:SPG-JR04,Muhanji:2019:SPG-JR05}. Typically, it includes simulation functionality for two energy market layers: the Security Constrained Unit Commitment (SCUC) and the Security Constrained Economic Dispatch (SCED), power system regulation and a physical model of the power grid itself (i.e. power flow analysis).  For this study, the simulator has been customized for ISO-NE operations to include the Real-Time Unit Commitment (RTUC) as shown in Fig.~\ref{fig:epecs}.  Furthermore, the SGEM model\cite{Rutberg:2011:00,Muhanji:2019:SPG-JR05} is used to capture the essential physics of cooling processes for thermal power plants and in turn compute the water withdrawals and consumption for each power plant.

\subsection{Regional Energy Deployment
System (ReEDS) for Capacity Planning}\label{sec:planning}
ReEDS is a capacity planning tool that was developed by the National Renewable Energy Laboratory (NREL) starting in 2003. ReEDS is a tool that identifies the long-term evolution of the electric power grid for various regions in the United States\cite{Cohen:2019:00,Short:2011:00,Eurek:2016:00}. At its core ReEDS is an optimization tool that identifies the cost-optimal mix of generation technologies subject to reliability, generation resource, and regulatory constraints\cite{Cohen:2019:00,Short:2011:00,Eurek:2016:00}. The optimization has a two-year time step for a total of 42 years ending in 2050\cite{Cohen:2019:00,Short:2011:00,Eurek:2016:00}. The final output of the simulation is generation capacity by technology, storage capacity, electricity costs among others\cite{Cohen:2019:00,Short:2011:00,Eurek:2016:00}. This optimization tool was used to determine the evolution of the ISO-NE system from the 2030 scenarios to the 2040 scenarios. The model input assumptions were selected from configurations defined by the 2018 Standard Scenarios\cite{Cole:2018:00} (see Table~\ref{tb:scens}) to align with the 2030 capacity mixes described in Section~\ref{sec:scenarios}. Details on added capacities for each scenario can be found in Section~\ref{sec:casestudy}.
\begin{table*}[!h]
\caption{ReEDS 2018 standard scenarios\cite{Cole:2018:00} used to evolve the SOARES 2030 scenarios into the 2040 scenarios.}
\label{tb:scens}
\begin{footnotesize}
\begin{center}
\begin{tabular}{lp{8cm}p{5cm}}\toprule
&\textbf{SOARES 2030 Scenarios}& \textbf{ReEDS Scenarios}\\\toprule
1& RPSs + Gas& High RE Cost \\\midrule
2& ISO Queue& Accelerated Nuclear Retirements \\\midrule
3&Renewables Plus& Low RE Cost                     \\\midrule
4& No Retirements beyond \newline Forward Capacity Auctions (FCA) \#10 & Low Wind Cost\\\midrule
5&ACPs + Gas& Extended Cost Recovery\\\midrule
6&Renewable Portfolio Standards (RPSs) +\newline Geodiverse Renewables& Low Natural Gas Prices\\\bottomrule         
\end{tabular}
\end{center}
\end{footnotesize}
\label{tab:scorecard}
\end{table*}

\subsection{The Security Constrained Unit Commitment (SCUC)}
The power system balancing operation commences with the day-ahead resource scheduling in form of the SCUC. It is performed the day before to determine the best set of generators that can meet the hourly demand at a minimum cost. The time step for the SCUC is 1-hour and it determines the optimal set of generators for the next 24-hours. A simplified version of this program is presented in \cite{Muhanji:2019:SPG-JR05} and the full version customized for ISO-NE operations is presented here\cite{Muzhikyan:2019:SPG-JR04}. Note that the SCUC formulation used for this study extends the methodology in \cite{Muzhikyan:2019:SPG-JR04} to also include ramping constraints for wind, solar, and hydro resources\cite{Muhanji:2019:SPG-JR05}.

\subsection{Real-Time Unit Commitment  (RTUC)}
The same day resource scheduling is conducted every hour through the RTUC. It uses an optimization program that is quite similar to that of SCUC but only commits and de-commits \emph{fast-start} units. Fast-start units are defined by their ability to go online and produce at full capacity within 15-30 minutes. The RTUC runs every hour with a 15-minute time step and a 4-hour look-ahead. The complete mathematics for the RTUC can be found in \cite{Muzhikyan:2019:SPG-JR04} with slight modifications to include ramping constraints for wind, solar, and hydro resources as presented in \cite{Muhanji:2019:SPG-JR05}.

\subsection{The Security Constrained Economic Dispatch (SCED)}
The real-time balancing operation is implemented through the SCED which is run every 10-minutes. The role of the SCED is to move available generator outputs to new set points in a cost-effective way. The SCED does not bring online any units but rather ramps up or down the available online units. The SCED methodology is presented in \cite{Muzhikyan:2019:SPG-JR04,Muhanji:2019:SPG-JR05} and similar to SCUC and RTUC, it has been extended to allow for the ramping of wind, solar, and hydro resources\cite{Muhanji:2019:SPG-JR05}.  A more comprehensive description of the EPECS methodology and mathematical formulations for each control layer can be found in \cite{Muzhikyan:2019:SPG-JR04,Muhanji:2019:SPG-JR05}. This methodology has been analyzed and validated by ISO-NE.

\subsection{Regulation}
A pseudo-steady-state approximation of the regulation service model that ties directly to a power flow model of the physical power grid is also used in this study. Normally, imbalances at the output of the regulation service would be represented in the form of frequency changes\cite{Gomez-Exposito:2011:02}. However, for steady-state simulations with 1-minute time step, the concept of frequency is not applicable. Instead, a designated \emph{virtual} swing bus consumes the mismatches between generation and load to make the steady state power flow equations solvable\cite{Muzhikyan:2019:SPG-JR04}. 

\subsection{Variable Renewable Energy}
Variable renewable energy resources in the EPECS simulator are studied as time-dependent, spatially distributed exogenous quantities that contribute directly to the net load. Where the term \emph{net load} here is defined as the difference between the aggregated system load and the total generation produced by VREs, tieline profiles and any transmission losses\cite{Muzhikyan:2019:SPG-JR04}. 

As previously defined in \cite{Muhanji:2019:SPG-JR05}, the EPECS simulator differentiates energy resources into several classes:  

\begin{defn}
\textbf{\emph{Variable Renewable Energy Resources (VREs)}}: Generation resources with a stochastic and intermittent power output. Wind, solar, run-of-river hydro, and tie-lines are assumed to be VREs.  
\end{defn}

\begin{defn}
\textbf{\emph{Semi-Dispatchable Resources}}: Energy resources that can be dispatched downwards (i.e curtailed) from their uncurtailed power injection value. When curtailment is allowed for VREs, they become semi-dispatchable.  In this study, wind, solar and tie-lines are treated as semi-dispatchable resources. Note that for the purposes of this study, run-of-river and conventional hydro resources can be curtailed and, therefore, are treated as semi-dispatchable in the ``flexible case" mentioned below. However, in the conventional case, run-of-river and conventional hydro resources are \emph{not} semi-dispatchable.
\end{defn}

\begin{defn}
\textbf{\emph{Must-Run Resources}}: Generation resources that must run at their maximum output at all times. In this study, nuclear generation units are assumed to be \emph{must run} resources.
\end{defn}

\begin{defn}
\textbf{\emph{Dispatchable Resources}}:  Energy resources that can be dispatched up and down from their current value of power injection. In this study, all other resources are assumed to be dispatchable.
\end{defn} 

The EPECS simulator employs the operating reserve concepts described in \cite{Holttinen:2012:00,Ela:2010:00} with only a few changes. This study focuses on the normal operating reserves that are able to respond to real-time changes in wind and solar generation. Specifically, how much of these reserve quantities comes from water resources such as conventional hydro, run-of-river hydro, and water and waste-water treatment facilities. Normal operating reserves are classified as load following, ramping, and regulation reserves based on the mechanisms upon which they are acquired and activated.   For the purposes of this study, the curtailment of VREs  was assumed to provide both load-following and ramping reserves in an upward direction to their forecasted value and in a downward direction to their minimum operating capacity limit.  

\begin{figure}[!t]
\centering
\includegraphics[width=6.2in]{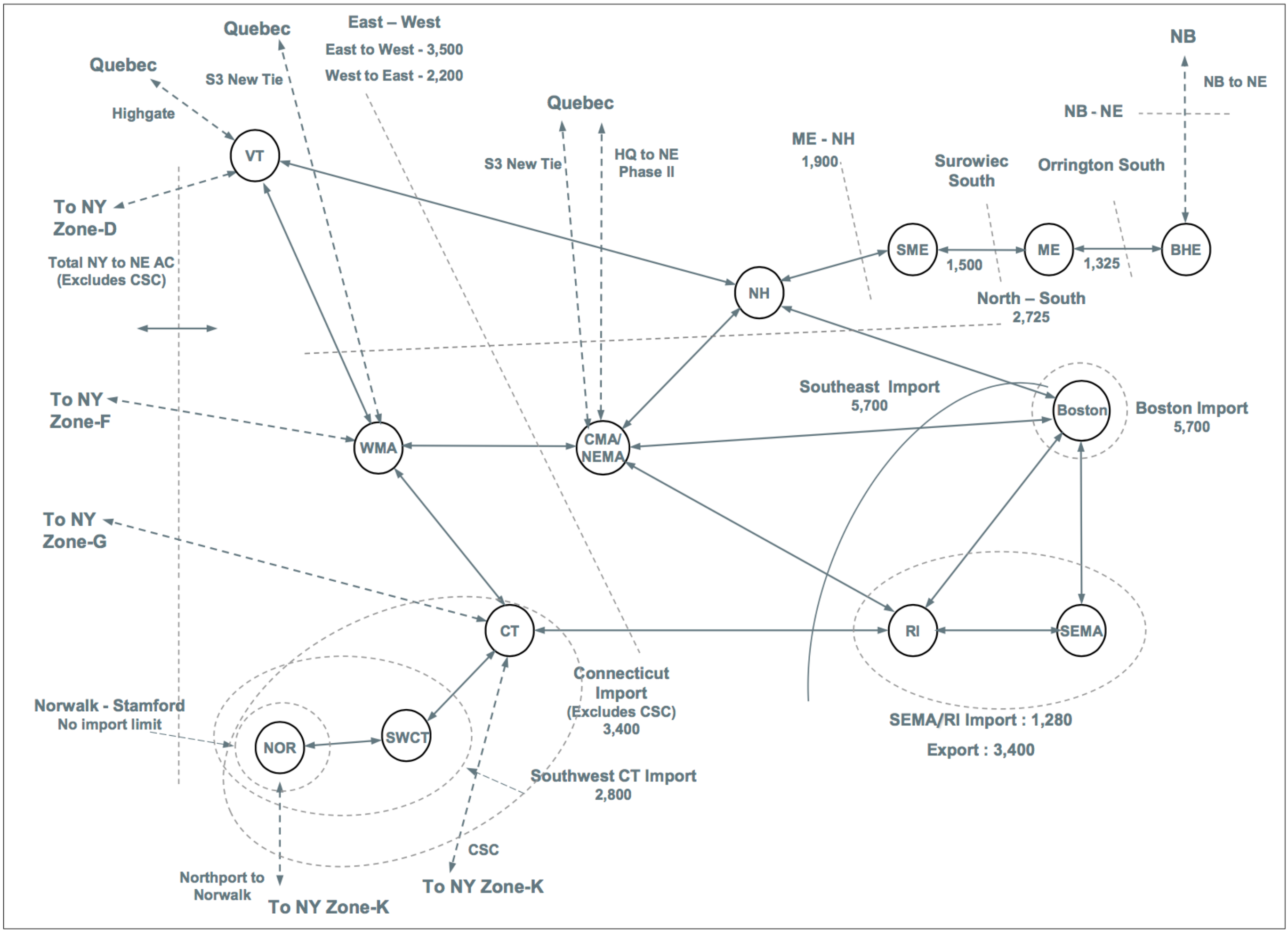}
\caption{ The ISO-NE zonal network model represented as ``pipes'' and ``bubbles''\cite{Muzhikyan:2019:SPG-JR04}}
\label{fig:pipenbubbles}
\end{figure}

These three types of operating reserves work together to respond to real-time forecast errors and variability in the \emph{net load} during normal system operation. Note that the actual quantities of these reserves are physical properties of the power system and exist regardless of whether they are monetized or not. The EPECS simulator provides as output quantities: system imbalances, operating reserves (load-following, ramping and regulation), generator set points, curtailed generation and line flows for every minute. 

\subsection{System-level Generic Model (SGEM)}
The S-GEM was developed to study the water use of fossil fuel,nuclear,geothermal and solar thermal power plants using either steam or combined cycle technologies\cite{Rutberg:2012:00,Integrated-Pollution-Prevention-and-Control-IPPC:2001:00,ECOFYS:2014:00,Eurelectric:1999:00,Tsou:2013:00,Pan:2018:01}. This model is also geography and case-independent; making it ideal for application to the ISO-NE system.  Three main cooling processes are applied in this paper: once-through cooling, wet tower cooling and dry-air cooling. Majority of the older generation power plants used once-through cooling technology while the newer power plants were either recirculating or dry-cooling. The formulae for computing water withdrawals and consumption are presented in \cite{Muhanji:2019:SPG-JR05}.

With this information, the energy-water flows through the system boundary of Fig.~\ref{fig:ewnflowchart} can be easily quantified (as detailed in \cite{Muhanji:2019:SPG-JR05}) to determine, water withdrawals and consumption by thermal power plants, as well as other aspects such as fuel consumption and $CO_2$ emissions. As illustrated in Figure~\ref{fig:ewnflowchart}, it is important to capture all the physical flows between the three physical infrastructures(water supply system, wastewater management system, and electricity supply system) and the natural surface environment. In this study, however, each water resource fits within an electric power system load area (or ```bubble" as they commonly called within the New England Power Pool).  Therefore, full hydraulic modeling does not provide additional insight in the provision of flexibility services to the electric power grid. The approach presented here is sufficient to capture all the interfaces between the water supply system and the electricity supply system and impose aggregate energy constraints as necessary. 



\section{Case Study Scenarios and Data}\label{sec:casestudy}
\begin{figure}[!htpb]
\centering
\caption{Summary of available generation capacity as a percentage of total available capacity by fuel type for all six 2040 scenarios.}
\includegraphics[width=6.4in]{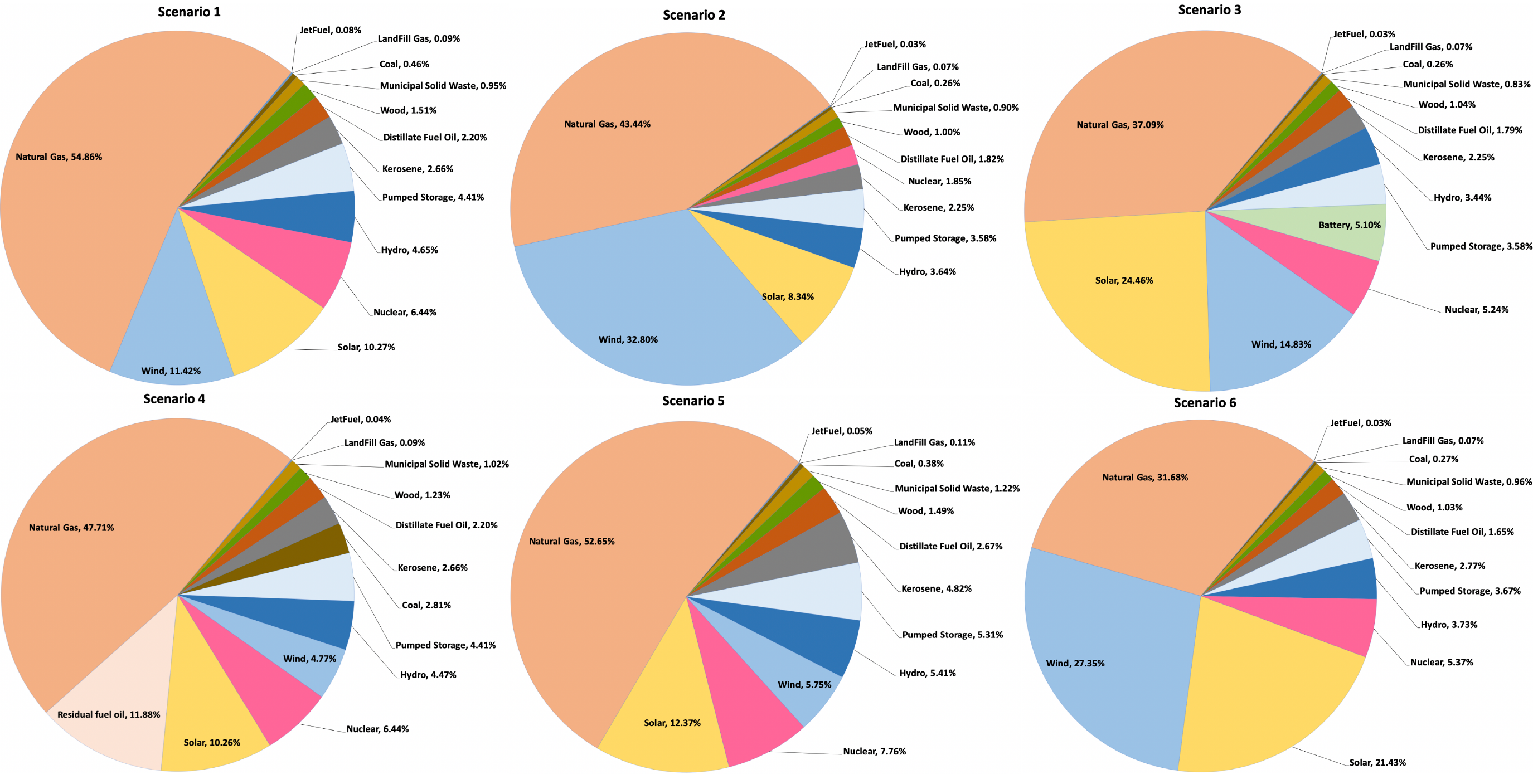}
\label{fig:scen_mixes}
\end{figure}
\subsection{Study Scenarios}\label{sec:scenarios}
The case study scenarios presented in this work are best understood in the context of the twelve scenarios that were studied in the 2017 System Operational
Analysis and Renewable Energy Integration Study (SOARES) that was commissioned by ISO-NE. These 12 scenarios distinguished between the amount and diversity of dispatchable generation resources, load profiles, and the penetration of VREs\cite{Muzhikyan:2019:SPG-JR04}. Of these scenarios, six were meant to describe 2025 while the other six were meant to describe 2030.  In the study presented here, the six 2030 (SOARES) scenarios were evolved forward ten years using the ReEDS capacity expansion software\cite{Cohen:2019:00,Short:2011:00,Eurek:2016:00,Cole:2018:00}. The final capacity mixes of the six scenarios are summarized in Figure~\ref{fig:scen_mixes} and are described further below. 


In order to assess the value of coordinated vs uncoordinated energy-water nexus operation, each of these six scenarios were simulated twice; once with energy-water resources as variable resources and another as semi-dispatchable resources.  These scenario variants are respectively referred to as the ``conventional" operating mode (as a control case) and the ``flexible" operating mode (as the experimental case).  
\subsubsection{\textbf{Scenario 1: RPSs + Gas}}\label{sec:scen1}

In this scenario, the oldest oil and coal generation units are retired by 2030 and the retired units are replaced by natural gas combined-cycle (NGCC) units at the same locations. Furthermore, the ReEDS model adds 50 MW of biomass, 233 MW of solar, 75MW of hydro and 6351 MW of natural gas (NG) to this scenario. It also retires 870 MW of nuclear, 667 MW of NG and 1127 MW of oil generation.

\subsubsection{\textbf{Scenario 2: ISO Queue}}\label{sec:scen2}
The retired oil and coal units from Scenario 1 are replaced by renewable energy resources instead of NGCC. The locations of the renewable energy resources are determined according to the ISO-NE Interconnection Queue. The ReEDS model resulted in the addition of 2498 MW of solar, 9.77 MW of hydro, and 5831.75 MW of NG (mostly in New Hampshire). In addition, 2471 MW of nuclear, 668 MW of natural gas and 25 MW of coal generation units were retired.

\subsubsection{\textbf{Scenario 3: Renewables Plus}}\label{sec:scen3}
In this scenario, more renewable energy resources are used to replace the retiring units. Additionally, battery energy systems, energy efficiency and plug-in hybrid electric vehicles (PHEV) are added to the system. Moreover, two new tie lines are added to increase the amounts of hydroelectricity imports. The ReEDS model results in the following modifications to this scenario: 1) addition of 2760 MW of solar, 9 MW of hydro, 2765 MW of NG, and 2) the retirement of 378 MW of coal, 870 MW nuclear, 667 MW of NG and 1127 MW of oil.

\subsubsection{\textbf{Scenario 4: No Retirements beyond Forward Capacity Auctions (FCA) \#10}}\label{sec:scen4}
In contrast to other scenarios, no generation units are retired beyond the known FCA resources. The FCA resources are replaced by NGCC located at the Hub. This scenario is the \emph{business-as-usual} scenario. The ReEDS model results in the following modifications to this scenario: 1) addition of 989 MW of solar, 4.2 MW of hydro, and 3987 MW of NG, and 2) the retirement of 383 MW of coal, 870 MW nuclear, 667 MW of NG and 1127 MW of oil.

\subsubsection{\textbf{Scenario 5: ACPs + Gas}}\label{sec:scen5}
In this scenario, the oldest oil and coal generation units are retired by 2030 and these units are replaced by new NGCC units to meet the net Installed Capacity Requirement (NICR). 
The ReEDS model results in the following modifications to this scenario: 1) addition of 3089 MW of solar, 11.1 MW of hydro, and 2496 MW of NG, and 2) the retirement of 253 MW of coal, 870 MW nuclear, 667 MW of NG and 1127 MW of oil.

\subsubsection{\textbf{Scenario 6: Renewable Portfolio Standards (RPSs) + Geodiverse Renewables}}\label{sec:scen6}
This scenario is similar to Scenario 5 but instead of replacing the retired units with NGCC units, additional renewable energy generation is used to meet the RPSs and the NICR. However, the solar PV and offshore wind units are located closer to the main load centers while the onshore wind is located in a remote area in Maine.
 The ReEDS model results in the following modifications to this scenario: 1) addition of 3011 MW of solar, 6.2 MW of hydro, and 2430 MW of NG, and 2) the retirement of 870 MW nuclear, 667 MW of NG and 1127 MW of oil.
 
The system data is consolidated into the zonal network model shown in Figure~\ref{fig:pipenbubbles}. The zonal network captures the power flows between pre-defined load zones (i.e. ``bubbles") along abstracted ``pipes"; thus eliminating the need for Critical Energy/Electric Infrastructure Information (CEII) clearance. The EPECS simulator implements a lossless DC Power Flow Analysis to determine these flows as described in \cite{Muzhikyan:2019:SPG-JR04,Muhanji:2019:SPG-JR05}. The high-level interface flow limits between the various bubbles are indicative of the line congestion often experienced in the ISO New England territory.  In addition to the changes in capacity mixes implemented in REEDS, interface limits were raised to reflect the likely situation that New England would work to resolve line congestion found in the 2025 and 2030 scenarios in the SOARES scenarios\cite{Muzhikyan:2019:SPG-JR04}.  Finally, in addition to the electric data, data on power consumption by water and waste-water treatment facilities as well as the cooling mechanisms of thermal generators were used to determine their share of the peak load.  The cooling data for thermal power plants was further enhanced by data from the Energy Information Agency's (EIA) databases\cite{EIA923:2019:00,EIA860:2019:00,EIA:2019:00}. 

\begin{figure}[!ht]
\centering
\includegraphics[width=6.2in]{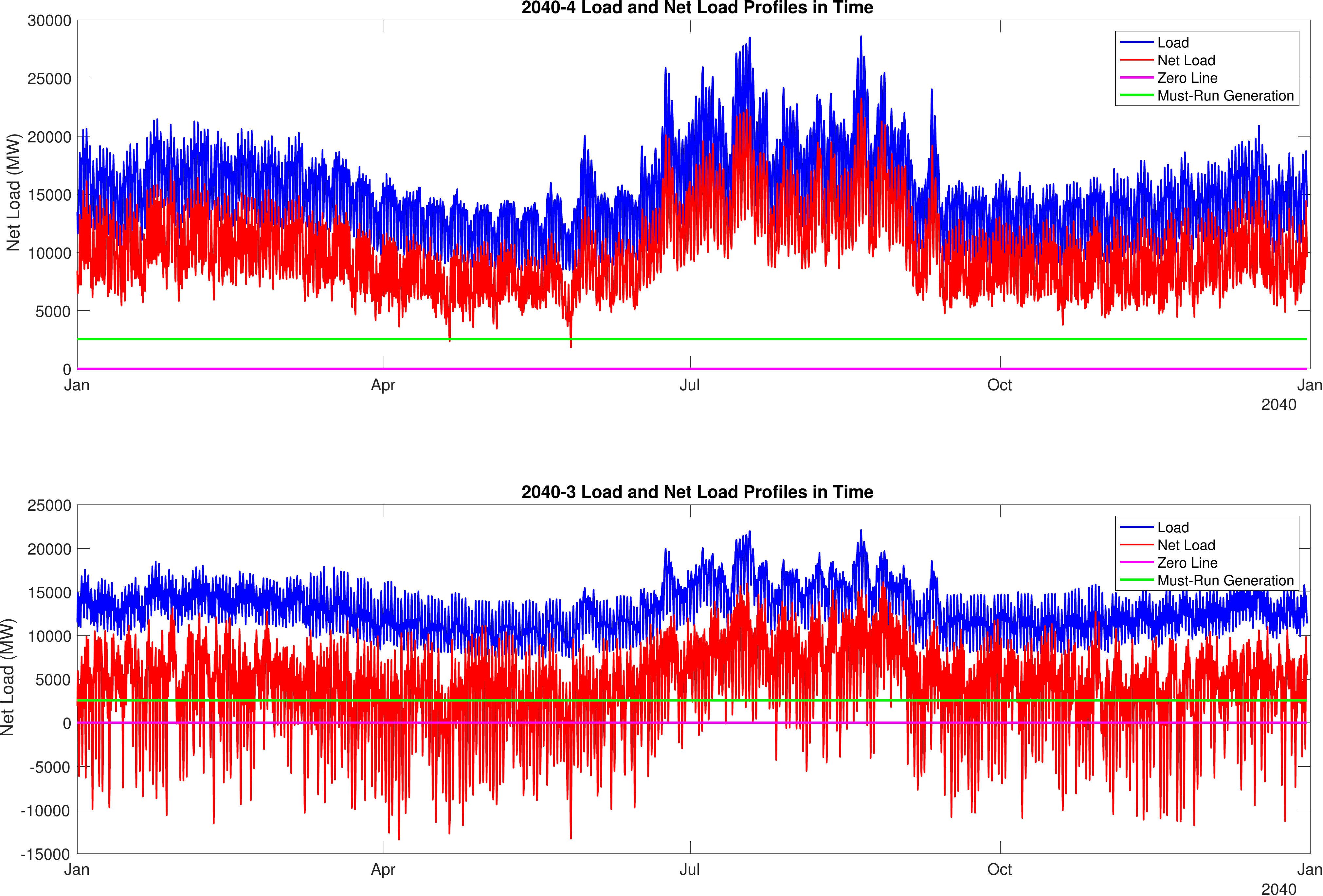}
\caption{The load and net load profiles from Scenario 2040-4 (top) and 2040-3 (bottom).}
\label{fig:204043NetLoad}
\end{figure}
\begin{figure}[!ht]
\centering
\includegraphics[width=6.2in]{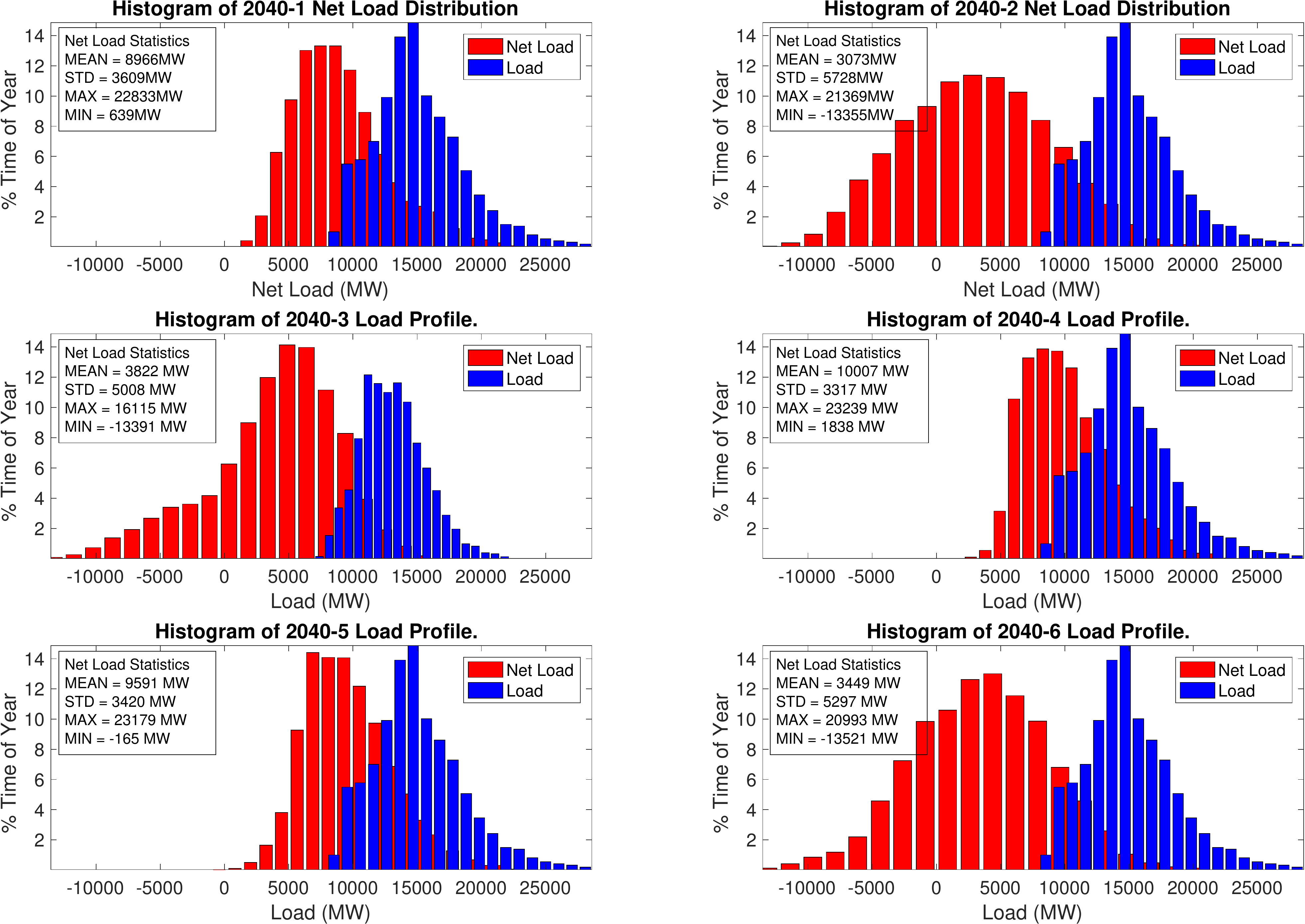}
\caption{A comparison of load and net load distributions for all six 2040 scenarios.}
\label{fig:2040NetLoadDist}
\end{figure}
\subsection{Net Load Profiles}
The net load profile comprised of the system load profile minus the wind, solar, tie-line, run-of-river and pond-hydro generation profiles. Figure \ref{fig:204043NetLoad} contrasts the net load profile of Scenario 2040-4 as a ``business-as-usual'' case to that of Scenario 2040-3 as a high VRE case.   The latter exhibits significant negative net load especially during low load periods such as the Spring and Fall seasons. Figure \ref{fig:2040NetLoadDist} summarizes the statistics of the net load profiles for all six scenarios.  The system \emph{peak load} for Scenarios 2040-1/2/4/5/6 was $28594 MW$ while that of Scenario 2040-3 was $22103 MW$ due to a higher penetration of energy efficiency measures. All scenarios had the same profile for electricity demand by water and wastewater treatment facilities. Run-of-river and pond-hydro generation profiles were curtailable at a price of $\$4.5/MWh$ similar to the 2017 ISO-NE SOARES. In this study, flexible water resources have load-shedding rather load-shifting capability and are assumed to contribute to operating reserves. The $709GWh$ of available pumped storage capacity is treated as dispatchable for all six scenarios throughout the study. Table~\ref{tb:netloadData} summarizes the capacity data for these flexible energy-water resources.  Again, in order to assess the "flexibility value" of these energy-water resources, each of the six scenarios is simulated in a conventional-uncoordinated mode of operation as well as a flexible-coordinated mode. 
\begin{table}[!htbp]
\caption{A summary of available flexible water resources in the system as percentage of the peak load.}
\includegraphics[width=6.2in]{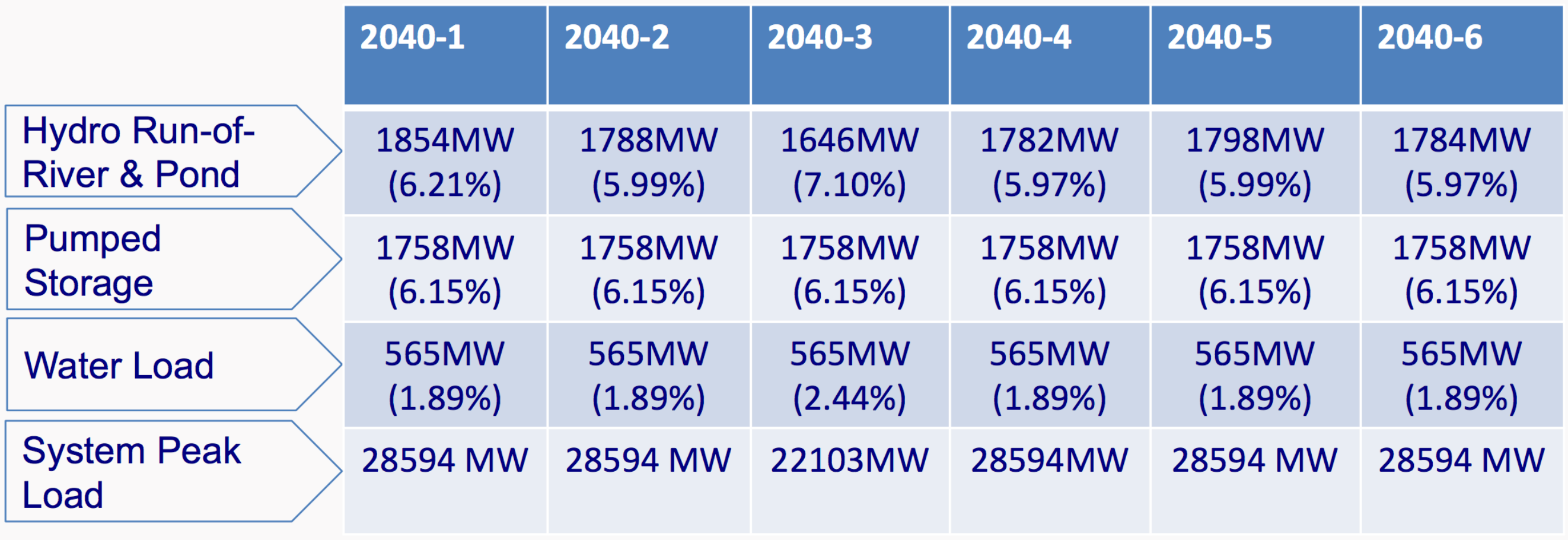}
\label{tb:netloadData}
\end{table}

\section{Results}\label{sec:results}
Given the aforementioned scenarios, the value of flexible energy-water resources is assessed from reliability, economic, and environmental perspectives.  From a reliability perspective, Section \ref{sec:balPerf} presents the relative improvements in the system's balancing performance  as quantified by the available quantities of operating reserves (i.e. load-following, ramping, and regulation reserves), curtailment, and the magnitude of system imbalances.  From an environmental perspective, Section \ref{sec:envImpact} quantifies the improvements in the quantities of water withdrawn and consumed as well as CO$_2$ emitted.  Here, \emph{water withdrawn} refers to the volumetric flow rate of water withdrawn from the natural surface environment and \emph{water consumption} refers to the amount of water not returned to its original point of withdrawal (due to evaporative losses).  Finally, Section \ref{sec:prodCosts} quantifies the associated production costs in the day-ahead and real-time energy markets.  

\subsection{Balancing Performance of Coordinated Energy-Water Operation}\label{sec:balPerf}
As mentioned above, this section presents the system balancing performance improvements as result of coordinated energy-water operation in terms of:  the available quantities of operating reserves (i.e. load-following, ramping, and regulation reserves), curtailment, and the magnitude of system imbalances.
\subsubsection{Load-Following Reserves}
In day-to-day operation, the upward and downward load-following reserves are used in time to allow the system to respond to variability and uncertainty in the net load. In the traditional operation of the electricity grid, having sufficient load-following reserves is a primary concern especially in systems with high penetrations of renewables. Both upward and downward load-following reserves are equally important in ensuring system reliability. As upward load following reserves are exhausted (approach zero), the ability of the system to respond to fluctuation in the net load is constrained.

\begin{figure}[!ht]
\centering
\includegraphics[width=6.2in]{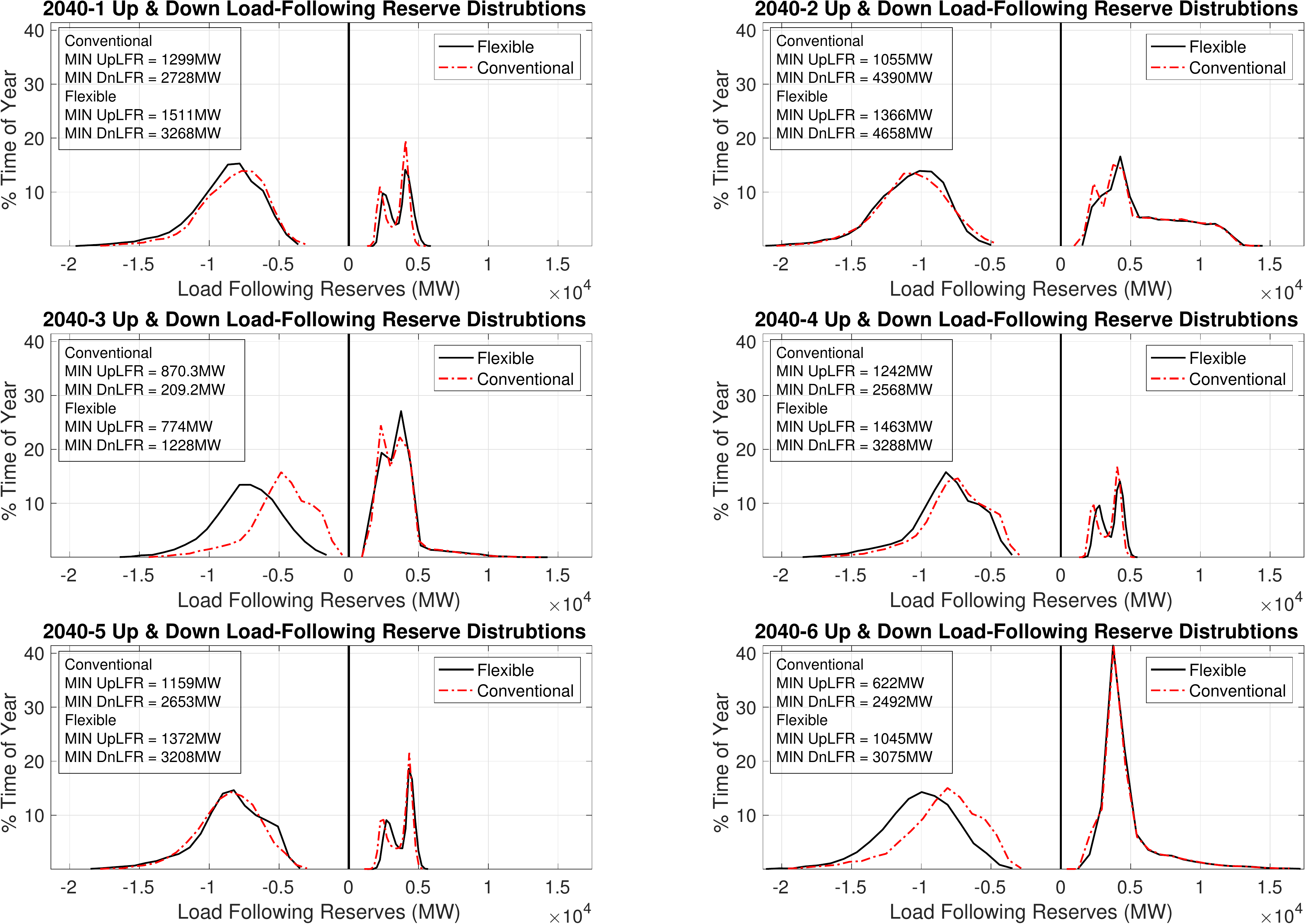}
\caption{Distributions of the available upward and downward load following reserves for all six 2040 scenarios in both the conventional and flexible operating modes.}
\label{fig:2040LFR}
\end{figure}
\begin{table}[!t]
\caption{Change in downward and upward load-following reserves statistics (flexible \textit{minus} conventional) for 2040 scenarios.}
\begin{footnotesize}
\begin{center}
\begin{tabular}{p{3.0cm}p{1.5cm}p{1.5cm}p{1.5cm}p{1.5cm}p{1.5cm}p{1.5cm}}\toprule
\textbf{$\Delta$ LFR (MW)} & \textbf{2040-1}& \textbf{2040-2}& \textbf{2040-3}& \textbf{2040-4}& \textbf{2040-5}& \textbf{2040-6}\\\toprule
\textbf{Up Mean}& 208.1 \newline(5.77\%)& 171.7 \newline(2.86\%)& 65.6 \newline(1.83\%)& 207.1 \newline(5.78\%)& 194.2 \newline(5.08\%)& 57.7 \newline(1.24\%)\\\midrule
\textbf{Up STD}&  8.4 \newline(1.00\%)& -55.6 \newline(-1.94\%)& -17.3 \newline(-1.22\%)& -42.1 \newline(-5.32\%)& -67.6 \newline(-8.36\%)& -36.09 \newline(-1.74\%)\\\midrule
\textbf{Up Max}& 178.3 \newline(3.07\%)& 228.3 \newline(1.56\%)& 335.3 \newline(2.32\%)& 242.5 \newline(4.37\%)& 107.9 \newline(1.92\%)& 686.8 \newline(3.94\%)\\\midrule
\textbf{Up Min}& 211.9 \newline(14.03\%)& 311.1 \newline(22.77\%)& -96.3 \newline(-12.45\%)& 221.2 \newline(15.12\%)& 212.6 \newline(15.50\%)& 422.6 \newline(40.46\%)\\\midrule
\textbf{Up 95 percentile$^1$}& 241.1 \newline(10.51\%)& 282.7 \newline(11.59\%)&  6.0 \newline(0.31\%)& 288.9 \newline(12.35\%)& 294.6 \newline(11.83\%)& 244.5 \newline(9.15\%)\\\midrule
\textbf{Down Mean}& 743.8 \newline(8.48\%)& 801.6 \newline(7.41\%)& 925.5 \newline(12.66\%)& 647.2 \newline(7.83\%)& 744.0 \newline(8.77\%)& 984.1 \newline(9.68\%)\\\midrule
\textbf{Down STD}& 8.75 \newline(0.36\%)& 16.29 \newline(0.66\%)& 36.01 \newline(1.52\%)& 2.98 \newline(0.12\%)& 9.50 \newline(0.39\%)& 67.97 \newline(2.55\%)\\\midrule
\textbf{Down Max} & 1177.0 \newline(6.11\%) & 932.5 \newline(4.37\%) & 1678.0 \newline(10.27\%) & 961.1 \newline(5.22\%) & 1086.0 \newline(5.79\%) & 1424.0 \newline(6.77\%) \\\midrule
\textbf{Down Min}& 540.3 \newline(16.53\%)& 267.9 \newline(5.75\%)& 1019.0 \newline(82.96\%)& 720.5 \newline(21.91\%)& 554.9 \newline(17.30\%)& 583.2 \newline(18.97\%)\\\midrule
\textbf{Down 95 percentile}& 749.0 \newline(13.96\%)& 790.6 \newline(10.79\%)& 1026.0 \newline(28.55\%)& 717.7 \newline(14.73\%)& 750.7 \newline(14.99\%)& 876.3 \newline(14.43\%)\\\bottomrule
\end{tabular}
\end{center}
\end{footnotesize}

\label{tab:LFRStats}
\end{table}

Therefore, an enhanced balancing performance with respect to load following reserves would show a significant trough around the zero LFR-axis in the distributions of load following reserves shown in Figure~\ref{fig:2040LFR}.  The larger the trough is, the more the system is not using its load following reserves to balance the system.  Figure~\ref{fig:2040LFR} shows that the flexible use of energy-water resources (in black) widens the trough of load-following reserves around the zero line relative to conventional operation (in red).  These graphical results are confirmed numerically in Table~\ref{tab:LFRStats}.  Flexible operation enhances the mean values of the upward and downward load following reserves (treated as separate distributions) by 1.24\%-- 12.66\% across all six scenarios.   Furthermore, the minimum upward and downward load following reserves are improved by flexible operation by 5.75\% -- 82.96\% across all but one of the six scenarios.  The minimum statistic is particularly important because it defines a type of worst case ``safety margin" that the system will always have available to ensure its security.  Similarly the 95 percentile statistic gives a measure of how much this minimum level increases when 5\% of the distribution is treated as abnormal outlier behavior.  The simulations show improvements in the 95 percentile statistic of 0.13--28.55\% across all six scenarios; thus demonstrating its robustness to not just the minimum worst-case point but also the distribution tail that represents challenging periods of operation.  The maximum and standard deviation statistics are provided for completeness.

\subsubsection{Ramping Reserves}
Ramping reserves describe the total amount of power that the system can respond up or down within a minute.  Traditionally, only dispatchable resources are assumed to contribute towards ramping reserves. In this study, renewable energy resources are semi-dispatchable by virtue of curtailment.  Consequently, they are assumed to not just be able to ramp down or up to their minimum or maximum values but also do so within five minutes given their power-electronics based control. Five minutes, in this case, coincides with the minimum time-step used in the real-time market. Similar to load-following reserves, ramping reserves are key to ensuring that the system can respond in time to fluctuations in the net load. Having sufficient amounts of both upward and downward ramping reserves is equally important to ensuring reliable performance. As the amount of ramping reserves approaches zero, the ability of the system to respond to net load variability is significantly diminished. 

\begin{figure}[!h]
\centering
\includegraphics[width=6.3in]{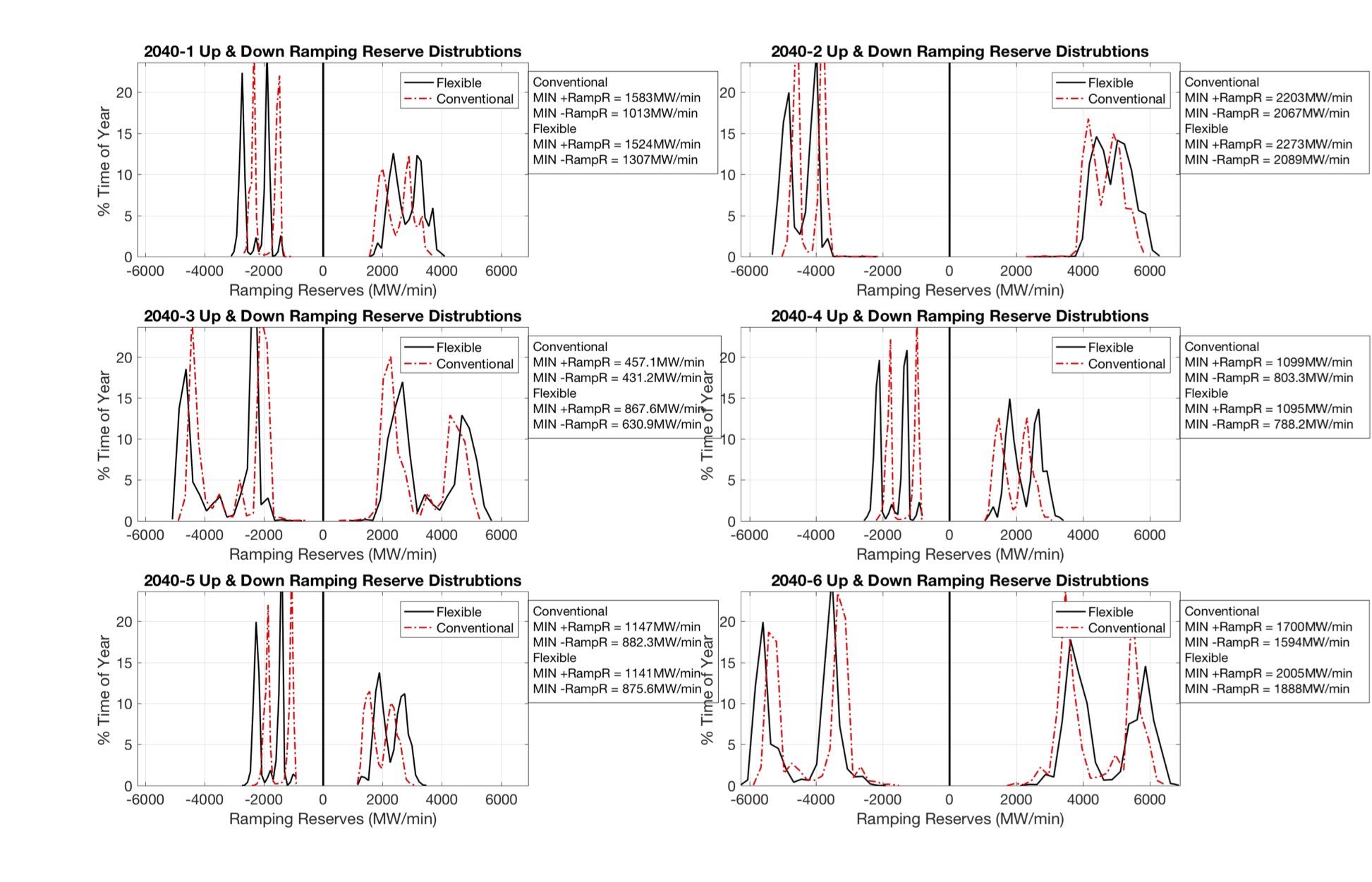}
\caption{Distributions of the available upward and downward ramping reserves for all six 2040 scenarios in both the flexible and conventional operating modes.}
\label{fig:2040RampR}
\end{figure}
\begin{table*}[!t]
\caption{Change in downward and upward ramping reserves statistics (flexible \textit{minus} conventional) for all six 2040 scenarios.}
\begin{footnotesize}
\begin{center}
\begin{tabular}{p{3.5cm}p{1.5cm}p{1.5cm}p{1.5cm}p{1.5cm}p{1.5cm}p{1.5cm}}\toprule
\textbf{$\Delta$ RampR Stats (MW/min)} & \textbf{2040-1}& \textbf{2040-2}& \textbf{2040-3}& \textbf{2040-4}& \textbf{2040-5}& \textbf{2040-6}\\\toprule
\textbf{Up Mean}& 334.9 \newline(11.83\%) & 259.4 \newline(5.28\%) & 291.3 \newline(8.26\%) & 308.7 \newline(13.78\%) & 325.3 \newline(14.31\%) & 287.7 \newline(6.16\%) \\\midrule
\textbf{Up STD}& 14.8 \newline(2.86\%) & 27.9 \newline(5.42\%) &  3.5 \newline(0.31\%) & 16.3 \newline(3.40\%) & 11.6 \newline(2.55\%) & 15.8 \newline(1.48\%) \\\midrule
\textbf{Up Max}& 430.7 \newline(10.40\%) & 354.7 \newline(5.65\%) & 271.0 \newline(4.83\%) & 361.5 \newline(10.43\%) & 372.9 \newline(10.58\%) & 331.1 \newline(4.79\%) \\\midrule
\textbf{Up Min}& -59.3 \newline(-3.89\%) & 69.7 \newline(3.07\%) & 410.6 \newline(47.32\%) & -4.4 \newline(-0.40\%) & -5.6 \newline(-0.49\%) & 305.1 \newline(15.21\%) \\\midrule
\textbf{Up 95 percentile}& 310.6 \newline(14.77\%) & 195.5 \newline(4.68\%) & 314.9 \newline(14.11\%) & 300.0 \newline(18.78\%) & 318.0 \newline(19.19\%) & 42.5 \newline(1.28\%) \\\midrule
\textbf{Down Mean}& 339.7 \newline(14.81\%) & 261.8 \newline(5.86\%) & 292.3 \newline(8.70\%) & 317.3 \newline(18.35\%) & 325.8 \newline(17.88\%) & 288.9 \newline(6.50\%) \\\midrule
\textbf{Down STD}& 16.4 \newline(3.69\%) & 21.4 \newline(4.81\%) &  1.5 \newline(0.13\%) & 16.1 \newline(3.67\%) & 12.7 \newline(2.94\%) & 12.4 \newline(1.20\%) \\\midrule
\textbf{Down Min}& 294.2 \newline(22.51\%) & 22.1 \newline(1.06\%) & 199.7 \newline(31.65\%) & -15.1 \newline(-1.92\%) & -6.7 \newline(-0.76\%) & 293.9 \newline(18.44\%) \\\midrule
\textbf{Down Max}& 417.3 \newline(15.37\%) & 354.3 \newline(7.06\%) & 275.9 \newline(5.64\%) & 385.1 \newline(17.38\%) & 345.1 \newline(14.42\%) & 320.7 \newline(5.40\%) \\\midrule
\textbf{Down 95 percentile}& 344.3 \newline(19.12\%) & 208.5 \newline(5.31\%) & 308.0 \newline(13.94\%) & 328.3 \newline(26.15\%) & 337.4 \newline(24.92\%) & 42.1 \newline(1.32\%) \\\bottomrule
\end{tabular}
\end{center}
\end{footnotesize}

\label{tab:RampStats}
\end{table*}

Similar to load-following reserves, both upward and downward ramping reserves are enhanced through the flexible operation of energy-water resources.  Figure~\ref{fig:2040RampR} illustrates a widened trough in the flexible operating mode relative to the conventional mode.  This observation is supported by the statistics in Table~\ref{tab:RampStats}. The mean value for the upward ramping reserves is improved across all scenarios by up to 14.31\%. Likewise, the mean downward ramping reserves are improved by up to 18.35\%. Another key measure of sufficient ramping reserves is the minimum level. As illustrated in Table~\ref{tab:RampStats}, flexible operation enhances the minimum downward ramping reserves by 31.65\% and the minimum upward ramping reserves by a maximum of 47.32\%. However, in cases with a lower penetration of VREs such as scenarios 2040-1/4/5, the minimum levels are slightly worse in the flexible case than in the conventional case.  Despite these anomalies, flexible operation improved 95\% percentile levels of upward and downward ramping reserves in all cases (by 1.28\%--26.15\%). These results show that the curtailment of VREs increases the flexibility to the system if they are used to provide ramping reserves. A complete summary of ramping reserves statistics for all six scenarios is found in Table~\ref{tab:RampStats}. 

\subsubsection{Curtailment}
\begin{figure}[!ht]
\centering
\includegraphics[width=6.2in]{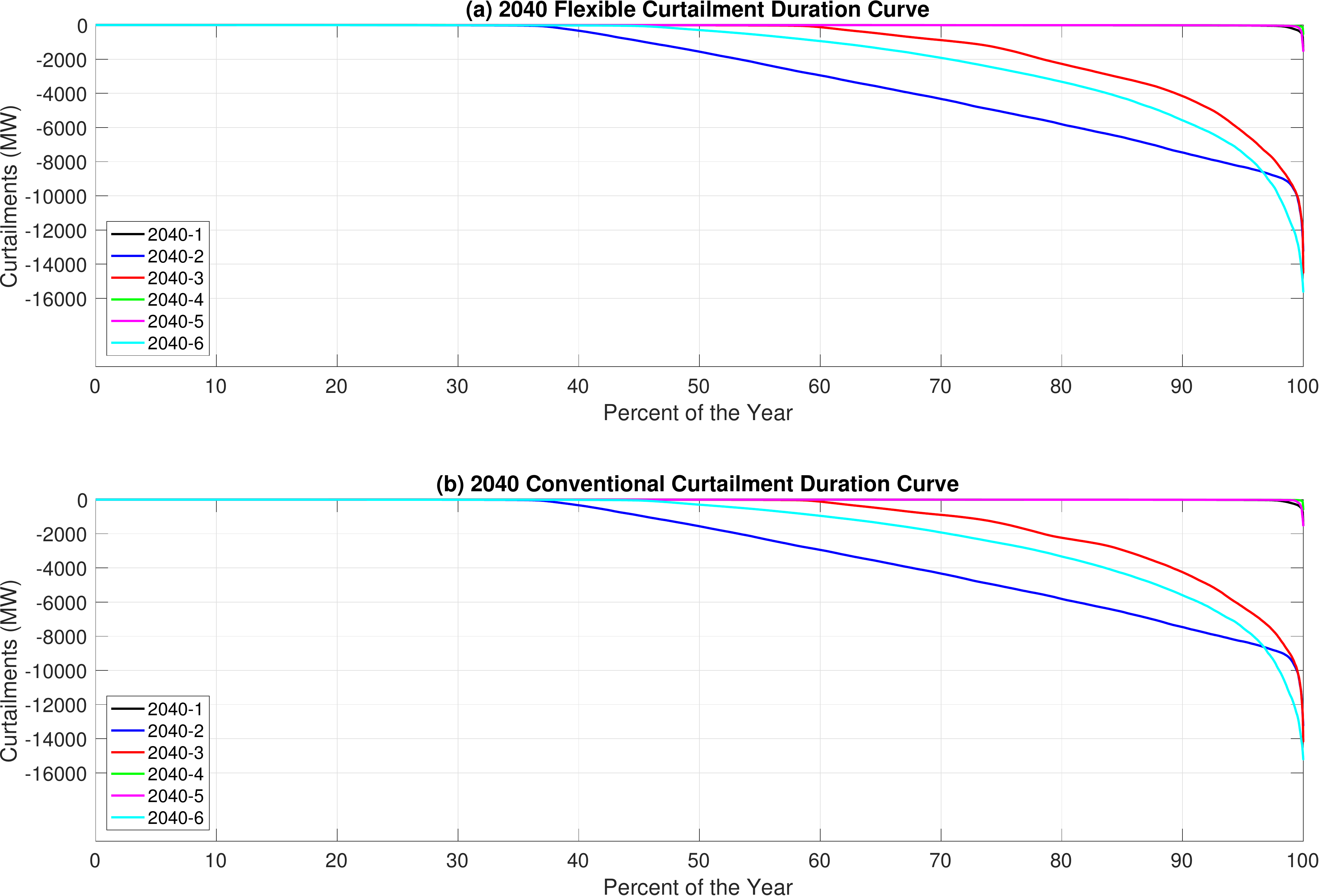}
\caption{Curtailment duration curves for all six 2040 scenarios in both the flexible (above) and conventional (below) operating modes.  }
\label{fig:Curtailment}
\end{figure}

By definition, flexible energy-water resources increase the amount of generation available for curtailment.  Recall that by Definition (2.6.2), run-of-river and conventional hydro-pond resources are semi-dispatchable resources that can be curtailed in a flexible operating mode.  As illustrated in Figure~\ref{fig:Curtailment}, scenarios with a lower penetration of VREs such as scenario 2040-1/4/5 curtail infrequently and the amount of megawatt curtailed is generally zero. For scenarios 2040-2/3/6, curtailment is used at least 40\% of the time. Although, the two case appear to have similar curtailment levels, a closer look at Table~\ref{tab:CurtStats} shows that the flexible case curtails for a smaller percentage of the year (2.67\% -- 10.9\%) less than the conventional case). Furthermore, the two operating modes show nearly identical levels of total curtailed energy.  In the absence of sufficient load-following and ramping reserves, curtailment serves a key role in ensuring system balance. This role is particularly crucial for VREs located in remote areas (e.g. Maine) where it serves as the only control option given topological constraints and distance from load areas.

\begin{table*}[!t]
\caption{Change in the curtailment statistics (flexible \textit{minus} conventional) for all six 2040 scenarios.}
\begin{footnotesize}
\begin{center}
\begin{tabular}{p{4.5cm}rrrrrr}\toprule
 & \textbf{2040-1}& \textbf{2040-2}& \textbf{2040-3}& \textbf{2040-4}& \textbf{2040-5}& \textbf{2040-6}\\\toprule
\textbf{Tot. Semi-Disp. Res.} (GWh)&  0.00 &  0.00 &  0.00 &  0.00 &  0.00 &  0.00 \\\midrule
\textbf{Tot. Curtailed Semi-Disp. \newline Energy} (GWh) & 17.71 & -1.95 & 60.86 & 23.44 & 20.57 & -6.18 \\\midrule
\textbf{\% Semi-Disp. Energy Curtailed}&  0.03 & -0.00 &  0.07 &  0.05 &  0.04 & -0.01 \\\midrule
\textbf{\% Time Curtailed}& -10.42 & -2.67 & -5.97 & -10.90 & -10.74 & -3.08 \\\midrule
\textbf{Max Curtailment Level} (MW)&  1.82 &  2.68 & 330.16 & -63.03 & -1.81 & 397.67 \\\bottomrule
\end{tabular}
\end{center}
\end{footnotesize}

\label{tab:CurtStats}
\end{table*}
\begin{figure}[!h]
\centering
\includegraphics[width=6.2in]{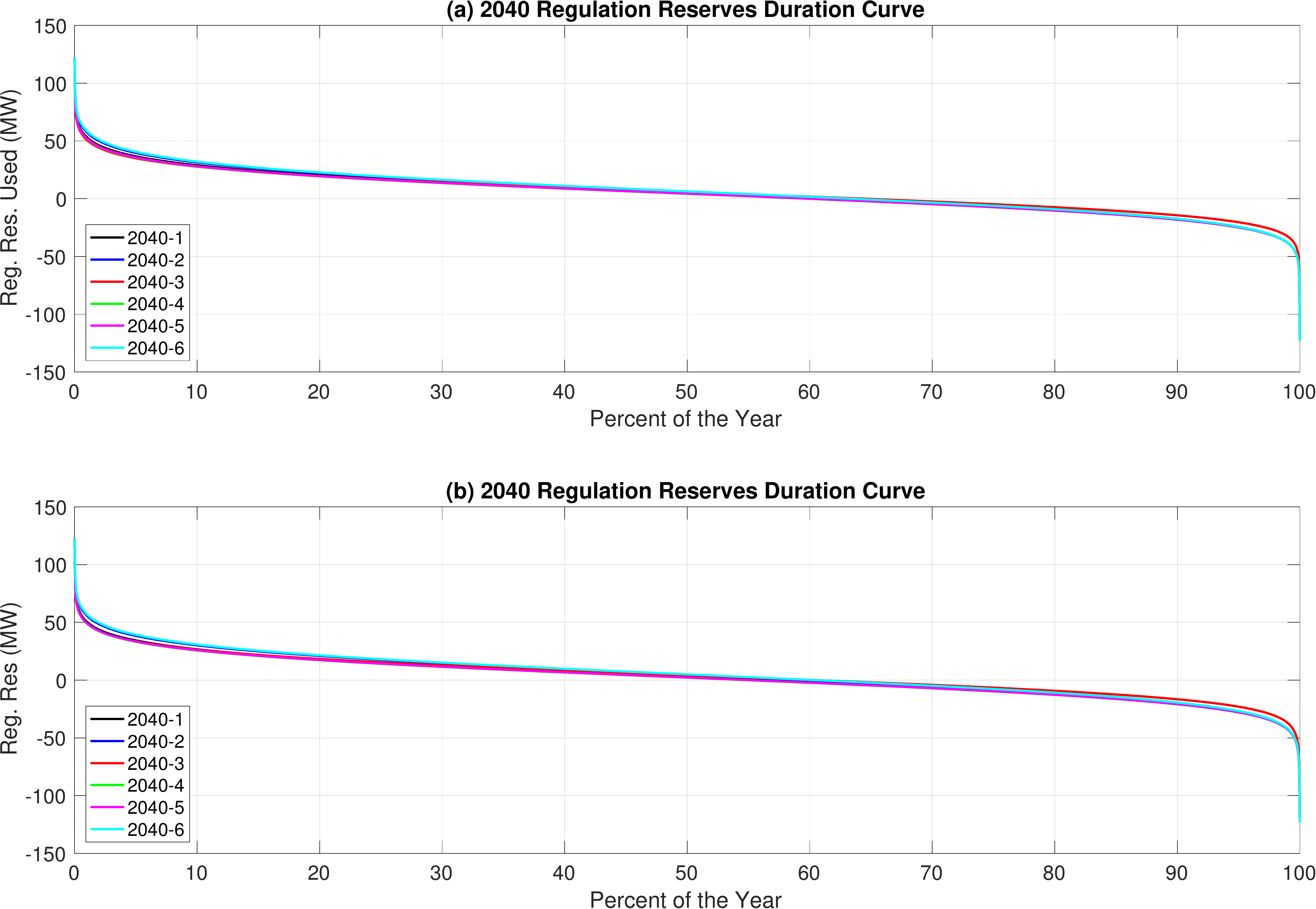}
\caption{Regulation duration curves for all six 2040 scenarios in both the flexible (above) and conventional (below) operating modes.}
\label{fig:regulation}
\end{figure}
\subsubsection{Regulation Service}
The regulation service is used to correct system imbalances in real-time. This control lever is used to meet any left-over imbalances after curtailment, load-following and ramping reserves have been used up during market operation. In both cases, all scenarios appear to use their regulation effectively as shown in Figure~\ref{fig:regulation}. This is indicative of a system that has sufficient regulation to mitigate real-time imbalances and maintain balancing performance. A closer inspection of Table~\ref{tab:RegStats} illustrates that flexible operation marginally increases the reliance on regulation (as shown by the excess mileage) and exhausts its regulation (albeit for a small fraction of the year ~0.001) for all but scenarios 2040-3 and 2040-4.


\begin{table*}[!h]
\caption{Change in regulation reserves statistics (flexible \textit{minus} conventional) for all six 2040 scenarios.}
\begin{footnotesize}
\begin{center}
\begin{tabular}{p{4.5cm}rrrrrr}\toprule
 & \textbf{2040-1}& \textbf{2040-2}& \textbf{2040-3}& \textbf{2040-4}& \textbf{2040-5}& \textbf{2040-6}\\\toprule
\textbf{\% Time Reg. Res \newline Exhausted}& 0.001 & 0.001 & 0.000 & 0.000 & 0.001 & 0.001 \\\midrule
\textbf{Reg. Res. \newline Mileage} (GWh)& 1.800 & 0.354 & 0.788 & 1.014 & 1.190 & 0.468 \\\bottomrule
\textbf{\% Reg. Res. \newline Mileage}& 1.349 & 0.251 & 0.638 & 0.777 & 0.909 & 0.326 \\\bottomrule
\end{tabular}
\end{center}
\end{footnotesize}

\label{tab:RegStats}
\end{table*}

\begin{figure}[!h]
\centering
\includegraphics[width=6.2in]{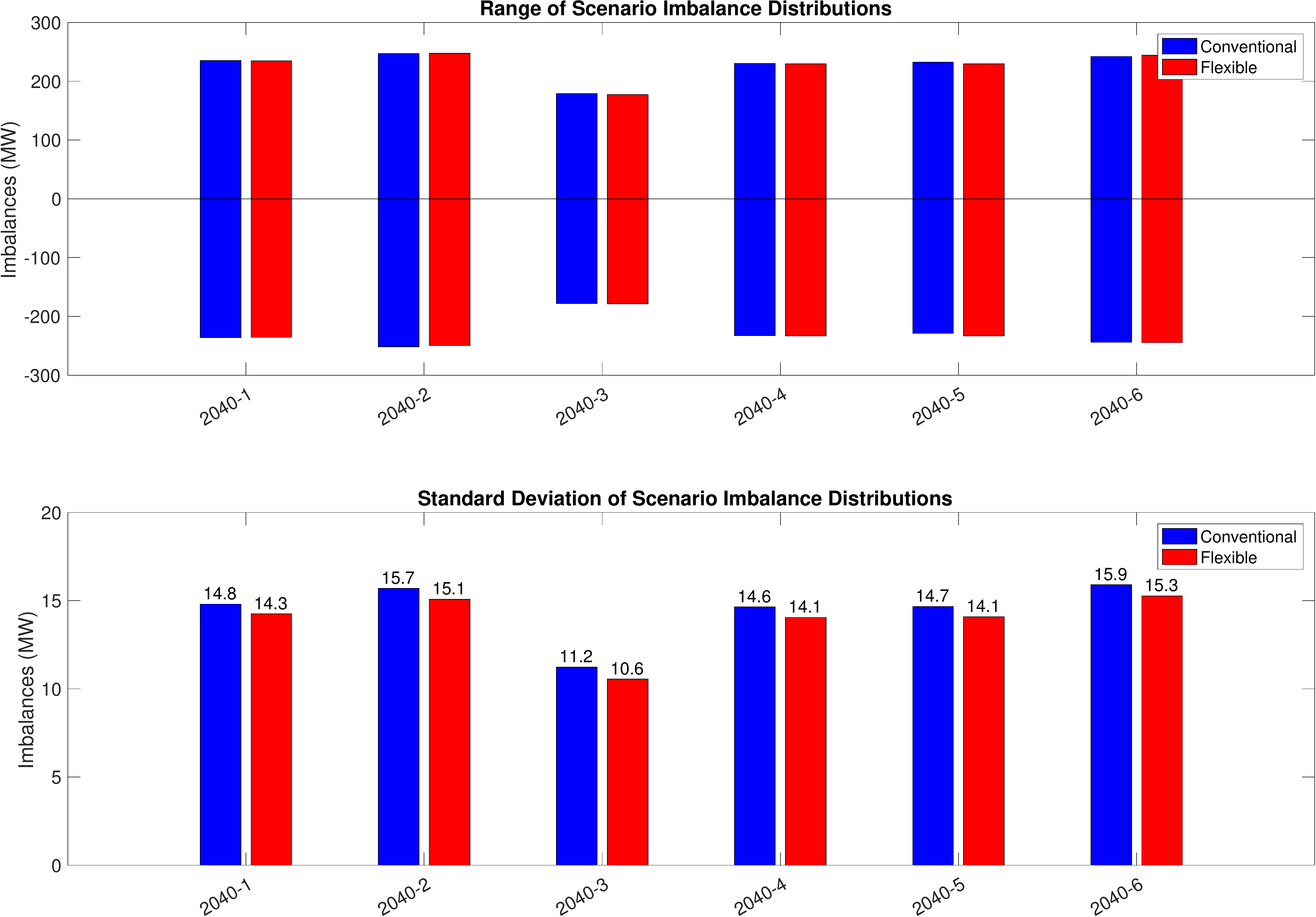}
\caption{Range (above) and standard deviation (below) statistics for all six 2050 scenarios in both the flexible (red) and conventional (blue) operation modes.  }
\label{fig:imbalances}
\end{figure}
\subsubsection{System Imbalances}

Balancing performance indicates the residual imbalances after the regulation service has been deployed. Given that the regulation service was barely saturated, the amount of imbalances are expected to be minimal. As shown in Figure~\ref{fig:imbalances}, flexible energy-water resources had a small impact on the range of final imbalances of the system. Both systems appear to perform similarly with all cases maintaining a standard deviation of less than 16MW across all six scenarios. Table~\ref{tab:ImbStats} illustrates that the flexible operating mode performs slightly better than the conventional with up to a 6.48\% improvement in standard deviation. The minimum imbalances are lower in all cases except for Scenarios 2040-1 and 2040-2. Similarly, the maximum imbalances are lower for the flexible operating mode except for Scenarios 2040-2 and 2040-6 which represent scenarios with high VREs.

\begin{table*}[!h]
\caption{Change in range and standard deviations of imbalances (flexible \textit{minus} conventional) for all six 2040 scenarios.}
\begin{footnotesize}
\begin{center}
\begin{tabular}{p{4.5cm}rrrrrr}\toprule
\textbf{Change in Imbalance} & \textbf{2040-1}& \textbf{2040-2}& \textbf{2040-3}& \textbf{2040-4}& \textbf{2040-5}& \textbf{2040-6}\\\toprule
\textbf{ Max (MW)}& -0.384 & 0.597 & -1.767 & -0.682 & -2.911 & 1.902 \\\midrule
\textbf{\% Max }& -0.164 & 0.241 & -0.998 & -0.297 & -1.269 & 0.779 \\\bottomrule
\textbf{ Min (MW)}& 0.118 & 1.831 & -0.598 & -0.363 & -4.405 & -0.462 \\\bottomrule
\textbf{\% Min }& -0.050 & -0.733 & 0.335 & 0.156 & 1.887 & 0.189 \\\midrule
\textbf{Std. (MW)}& -0.552 & -0.611 & -0.684 & -0.589 & -0.584 & -0.634 \\\midrule
\textbf{\% Std. }& -3.874 & -4.052 & -6.484 & -4.188 & -4.147 & -4.155 \\\bottomrule
\end{tabular}
\end{center}
\end{footnotesize}

\label{tab:ImbStats}
\end{table*}

\newpage 
\subsection{Environmental Performance of Coordinated Energy-Water Operation}\label{sec:envImpact}
As mentioned before, the environmental performance of coordinated energy-water operation is assessed through overall reductions in water withdrawals, consumption and CO$_2$ emissions.
\subsubsection{Water Withdrawals}
\begin{figure}[!h]
\centering
\includegraphics[width=6.2in]{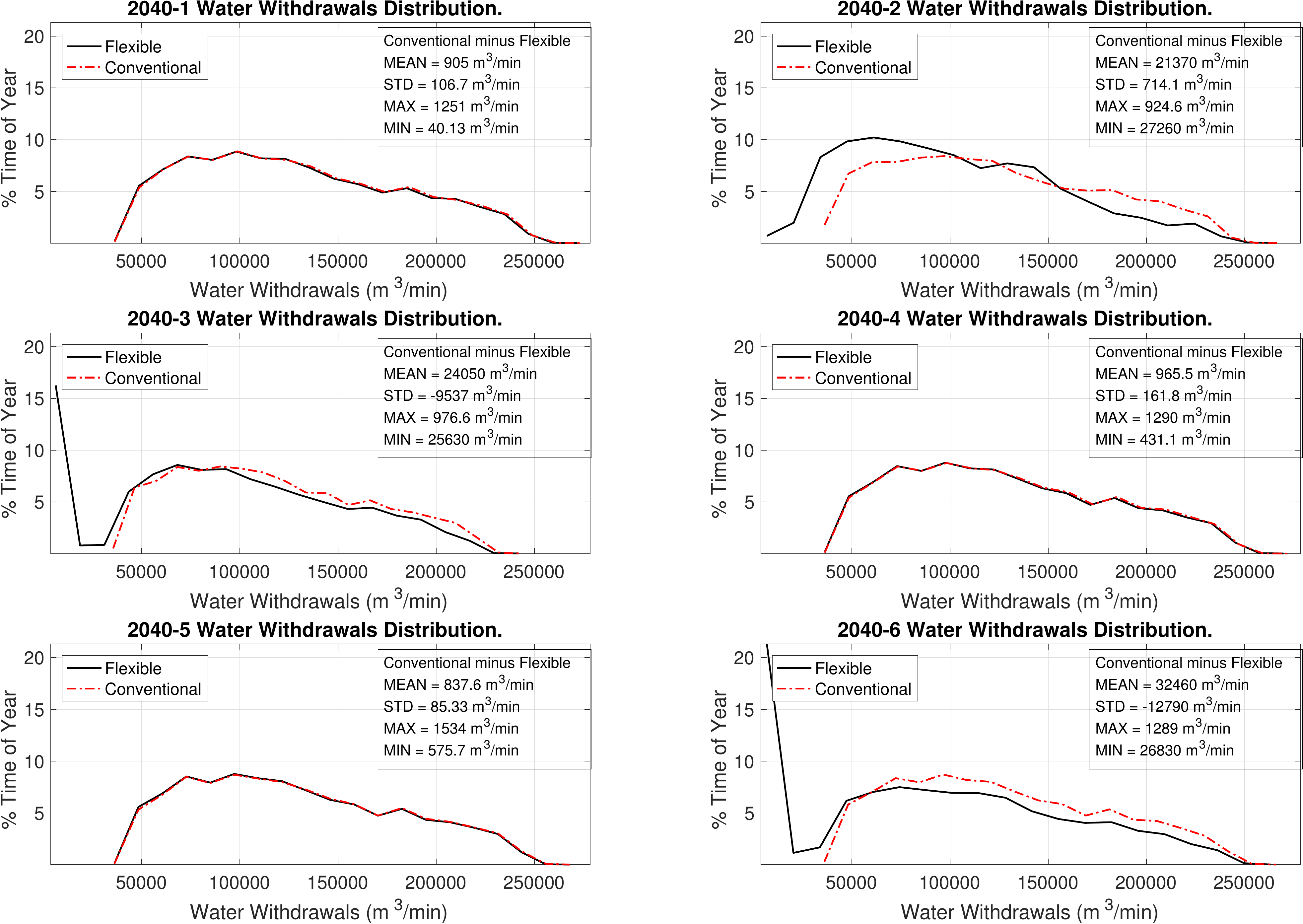}
\caption{Distributions of water withdrawals for all six 2040 scenarios in both the flexible and conventional operating modes.}
\label{fig:H20Withdrawals}
\end{figure}

Figure~\ref{fig:H20Withdrawals} shows the water withdrawal distributions for the flexible and conventional operating modes.  Flexible operation results in significantly lower withdrawals compared to conventional operation because the flexible energy-water resources are able to offset the use of thermo-electric power plants in favor of VREs. This phenomena is seen in how the flexible withdrawal distributions are shifted left towards zero. The associate water withdrawal statistics are summarized in Table~\ref{tab:H2OWStats} indicating improvements in mean withdrawals of up to 25.58\%. These improvements are most pronounced in Scenarios 2040-2/3/6 with high penetrations of VREs.  Indeed, the integration of several percent (on capacity basis) of flexible energy-water resources as shown in Table~\ref{tb:netloadData}, serves to reduce water withdrawals by many multiples of that percentage.  Such a phenomena can potentially appear in any scenario where VRE curtailment serves as a major lever of balancing control.  Nevertheless, the flexible operation of energy-water resources reduces water withdrawals  across all six scenarios.
\begin{table*}[!h]
\caption{Change in water withdrawals statistics (conventional \textit{minus} flexible) for all six 2040 scenarios.}
\begin{footnotesize}
\begin{center}
\begin{tabular}{p{4.1cm}p{1.5cm}p{1.5cm}p{1.5cm}p{1.5cm}p{1.5cm}p{1.5cm}}\toprule
\textbf{$\Delta H_20$ Withdrawals} & \textbf{2040-1}& \textbf{2040-2}& \textbf{2040-3}& \textbf{2040-4}& \textbf{2040-5}& \textbf{2040-6}\\\toprule
\textbf{ Mean} ($m^3/min$)& 905.0 \newline(0.70\%)& 21370.0 \newline(17.29\%)& 24050.0 \newline(20.59\%)& 965.5 \newline(0.74\%)& 837.6 \newline(0.65\%)& 32460.0 \newline(25.58\%)\\\midrule
\textbf{ STD} ($m^3/min$) & 106.7 \newline(0.20\%)& 714.1 \newline(1.35\%)& -9537.0 \newline(-19.92\%)& 161.8 \newline(0.31\%)&  85.3 \newline(0.16\%)& -12790.0 \newline(-24.40\%)\\\midrule
\textbf{ Max} ($m^3/min$)& 1251.0 \newline(0.45\%)& 924.6 \newline(0.34\%)& 976.6 \newline(0.39\%)& 1290.0 \newline(0.47\%)& 1534.0 \newline(0.56\%)& 1289.0 \newline(0.47\%)\\\midrule
\textbf{ Min} ($m^3/min$)&  40.1 \newline(0.11\%)& 27260.0 \newline(88.22\%)& 25630.0 \newline(75.82\%)& 431.1 \newline(1.17\%)& 575.7 \newline(1.54\%)& 26830.0 \newline(75.99\%)\\\midrule
\textbf{Total} ($m^3/min\times 10^6$)& 475.7 & 11230.0 & 12640.0 & 507.5 & 440.2 & 18090.0 \\\midrule
\textbf{Percent change} ($\%$)&  0.70 & 17.29 & 20.59 &  0.74 &  0.65 & 25.58 \\\bottomrule
\end{tabular}
\end{center}
\end{footnotesize}

\label{tab:H2OWStats}
\end{table*}

\subsubsection{Water Consumption}
\begin{figure}[!h]
\centering
\includegraphics[width=6.2in]{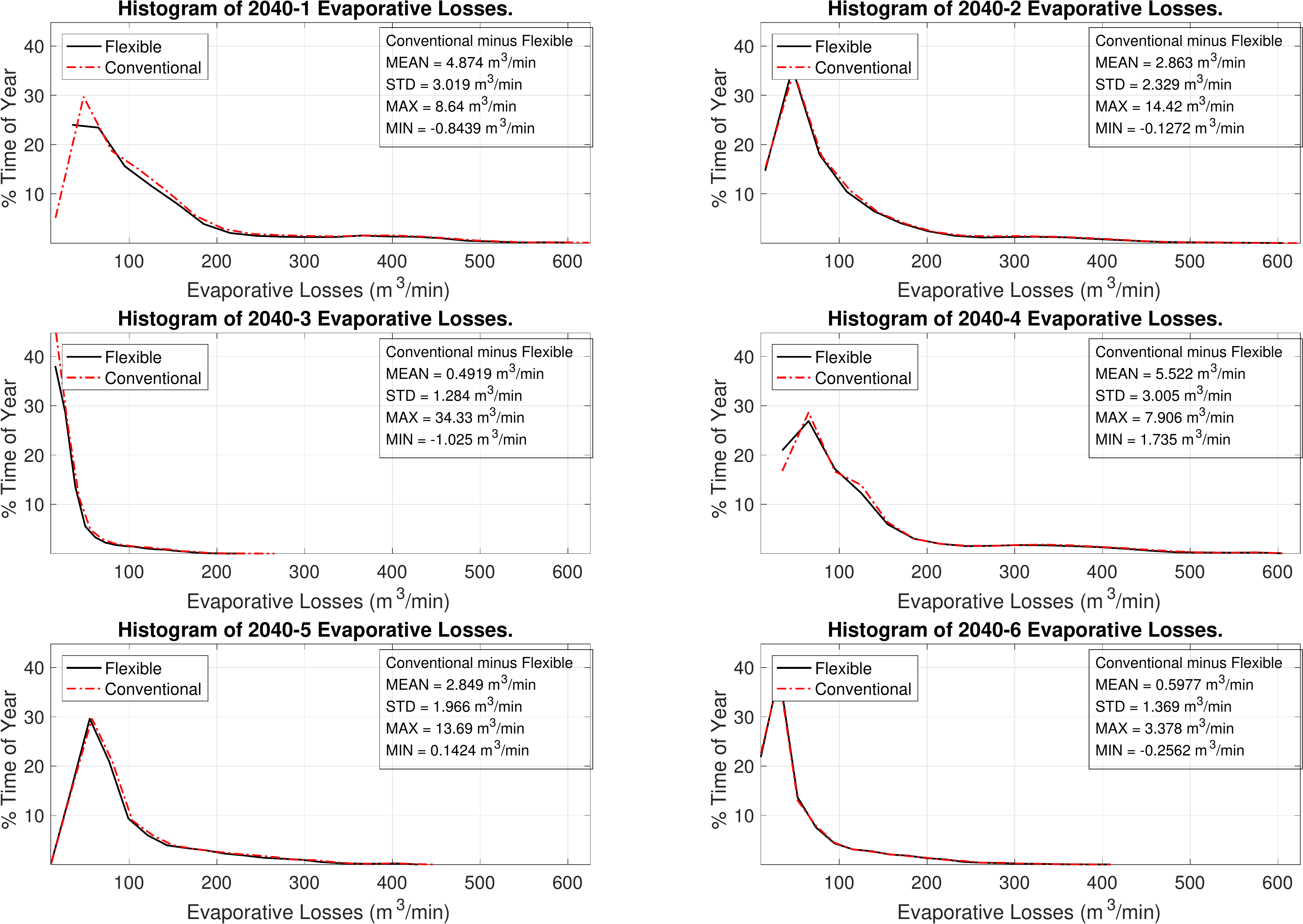}
\caption{Distributions of water consumption for all six 2040 scenarios in both the flexible and conventional operating modes.}
\label{fig:EvapDist}
\end{figure}

Electric power system water consumption occurs through the evaporative losses from cooling towers in recirculating cooling systems. Figure~\ref{fig:EvapDist} shows the water consumption distribution for both the conventional and flexible operating modes.  While the effect is not large, the flexible mode of operation shifts the distribution slightly towards the zero mark.   Specifically, flexible operation consumes 1.07--4.51\% less water than the conventional operation across all six scenarios, as shown in Table~\ref{tab:EvapStats}. This relatively small percentage nevertheless accounts for $258\times 10^3 m^3$ of water saved every year. Scenarios 2040-3 and 2040-6 have the least savings. Due to high penetrations of VREs, these scenarios require faster ramping generation which mostly comes from fast-ramping natural gas units with recirculating cooling systems. In short, the water saving effect of integrating VREs is a diminished to a certain extent by the need for operating reserves from water-consuming but flexible NGCC plants.  If demand side resources (from water loads or otherwise) played a large balancing role, then the water saving role of integrating VREs would be more pronounced.  

\begin{table*}[!t]
\caption{Change in evaporative loss statistics  (conventional \textit{minus} flexible) for all six 2040 scenarios.}
\begin{footnotesize}
\begin{center}
\begin{tabular}{p{4.0cm}p{1.5cm}p{1.5cm}p{1.5cm}p{1.5cm}p{1.5cm}p{1.5cm}}\toprule
\textbf{$\Delta$ Evap Losses}& \textbf{2040-1}& \textbf{2040-2}& \textbf{2040-3}& \textbf{2040-4}& \textbf{2040-5}& \textbf{2040-6}\\\toprule
\textbf{Mean} ($m^3/min$)&  2.67 \newline(3.96\%)&  1.63 \newline(3.11\%)&  0.30 \newline(1.44\%)&  3.37 \newline(5.03\%)&  1.51 \newline(2.84\%)&  0.31 \newline(1.03\%)\\\midrule
\textbf{STD} ($m^3/min$) &  1.10 \newline(2.77\%)&  1.05 \newline(2.97\%)&  0.74 \newline(5.58\%)&  1.23 \newline(3.33\%)&  0.61 \newline(2.61\%)&  0.68 \newline(3.05\%)\\\midrule
\textbf{Max} ($m^3/min$)&  5.71 \newline(2.45\%)&  3.42 \newline(1.44\%)&  6.40 \newline(6.02\%)& -0.00 \newline(-0.00\%)&  1.80 \newline(1.11\%)&  0.07 \newline(0.04\%)\\\midrule
\textbf{Min} ($m^3/min$)& -0.62 \newline(-3.50\%)& -0.00 \newline(-0.00\%)& -0.13 \newline(-1.65\%)&  0.47 \newline(2.56\%)& -0.12 \newline(-0.83\%)& -0.06 \newline(-0.52\%)\\\midrule
\textbf{Total} ($m^3\times 10^3$)&  1402 &   859 &   158 &  1769 &   794 &   165 \\\bottomrule
\textbf{Percent change} ($\%$)&  4.12 &  3.21 &  1.46 &  5.30 &  2.92 &  1.03 \\\bottomrule
\end{tabular}
\end{center}
\end{footnotesize}

\label{tab:EvapStats}
\end{table*}
\newpage 
\subsubsection{$CO_2$ Emissions}
\begin{figure}[!h]
\centering
\includegraphics[width=6.2in]{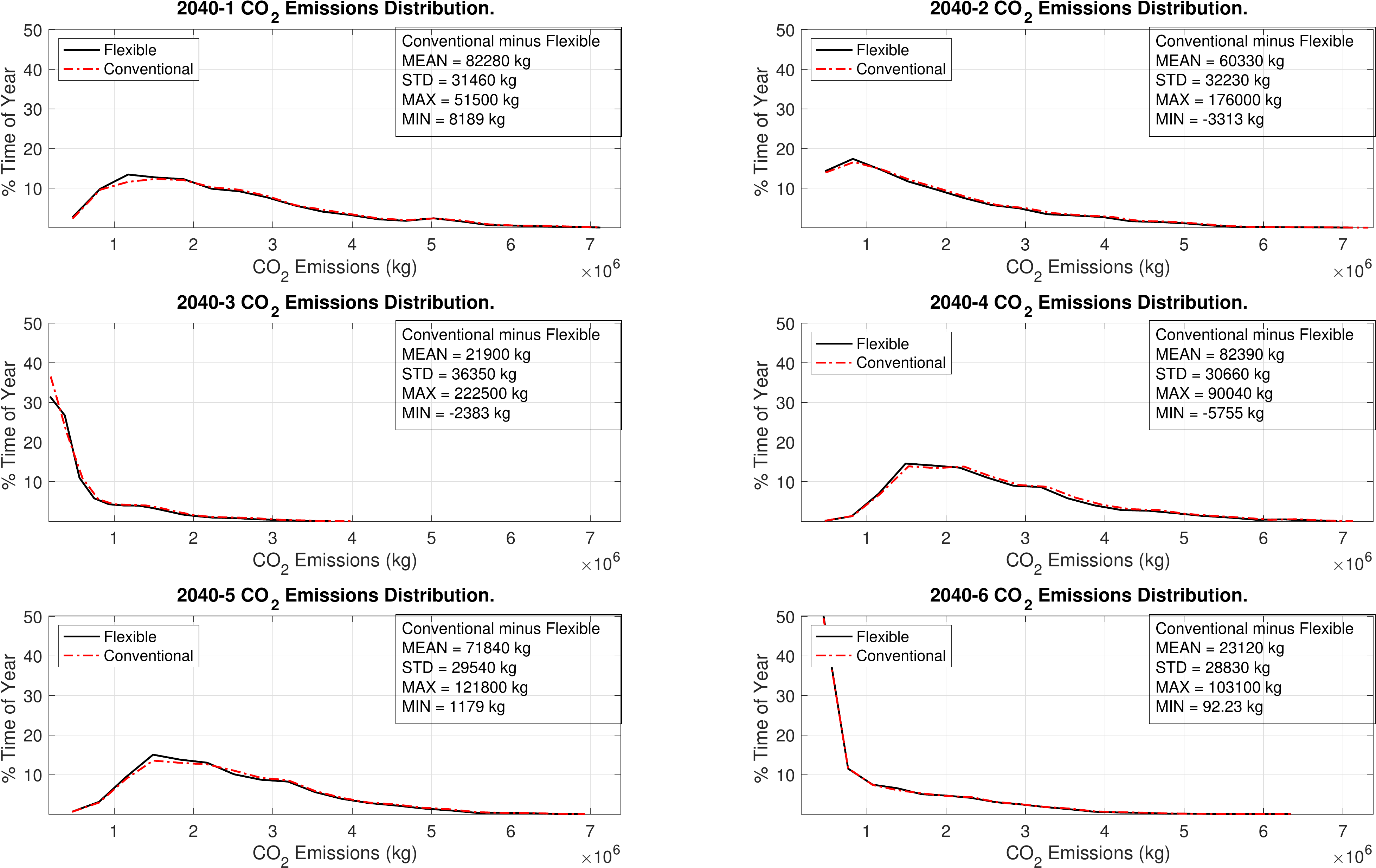}
\caption{Distributions of $CO_2$ emissions for all six 2040 scenarios in both the flexible and conventional operating modes.}
\label{fig:CDist}
\end{figure}

Finally, as shown in Figure~\ref{fig:CDist}, the overall $CO_2$ emissions are significantly reduced through flexible operation. It reduces the overall $CO_2$ emissions by 2.10\%--3.46\%, as shown in Table~\ref{tab:CO2Stats}. The mean, max, and standard deviation of emissions are all improved.  This CO$_2$ emissions reduction occurs because flexible energy-water resources 1.) eliminate the need for some generation through reduced electricity consumption, 2.) enable greater VRE generation through a reduction in curtailment and 3.) displace fossil-fueled conventional generation.  
\begin{table*}[!h]
\caption{Change in $CO_2$ emissions statistics (flexible \textit{minus} conventional) for all six 2040 scenarios.}
\begin{footnotesize}
\begin{center}
\begin{tabular}{p{4.1cm}p{1.5cm}p{1.5cm}p{1.5cm}p{1.5cm}p{1.5cm}p{1.5cm}}\toprule
$\Delta CO_2$ Emissions & \textbf{2040-1}& \textbf{2040-2}& \textbf{2040-3}& \textbf{2040-4}& \textbf{2040-5}& \textbf{2040-6}\\\toprule
\textbf{ Mean} ($kg$)& 82280 \newline(3.46\%)& 60330 \newline(3.28\%)& 21900 \newline(3.17\%)& 82390 \newline(3.11\%)& 71840 \newline(2.90\%)& 23120 \newline(2.10\%)\\\midrule
\textbf{ STD} ($kg$) & 31460.0 \newline(2.44\%)& 32230.0 \newline(2.66\%)& 36350.0 \newline(5.75\%)& 30660.0 \newline(2.69\%)& 29540.0 \newline(2.71\%)& 28830 \newline(2.96\%)\\\midrule
\textbf{Max} ($kg$)& 51500 \newline(0.71\%) & 176000 \newline(2.38\%) & 222500 \newline(5.54\%) & 90040 \newline(1.26\%) & 121800 \newline(1.72\%) & 103100\newline(1.59\%) \\\midrule
\textbf{ Min} ($kg$)& 8189.00 \newline(2.07\%)& -3313.00 \newline(-1.08\%)& -2383.00 \newline(-1.35\%)& -5755.00 \newline(-1.14\%)& 1179.00 \newline(0.31\%)& 92.23 \newline(0.03\%)\\\midrule
\textbf{Total} ($kg\times 10^6$)& 43240 & 31710 & 11510 & 43300 & 37760 & 12150 \\\midrule
\textbf{Percent change} ($\%$)&  3.46 &  3.28 &  3.17 &  3.11 &  2.90 &  2.10 \\\bottomrule
\end{tabular}
\end{center}
\end{footnotesize}

\label{tab:CO2Stats}
\end{table*}
\newpage 

\subsection{Economic Performance of Coordinated Energy-Water Operation}\label{sec:prodCosts}
The economic performance of coordinated energy-water operation is assessed in terms of the day-ahead and real-time production costs.
\subsubsection{Day-Ahead Energy Market Production Costs}
\begin{figure}[!h]
\centering
\includegraphics[width=6.2in]{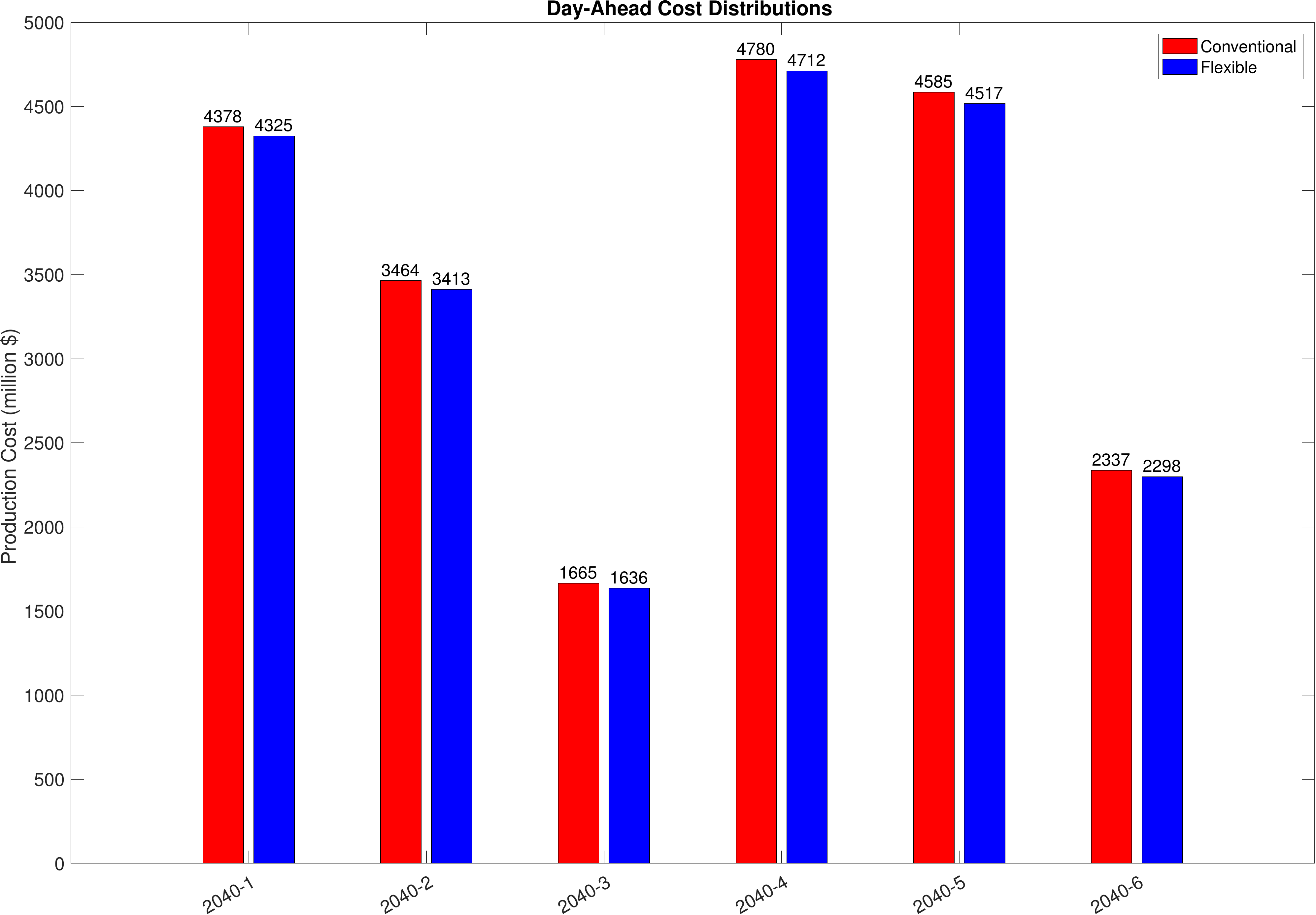}
\caption{Total production cost in the day-ahead energy market for all 2040 scenarios in both the flexible and conventional operating modes.  }
\label{fig:DACost}
\end{figure}

Figure~\ref{fig:DACost} shows flexible operation reduced the total production cost in the day-ahead energy market for all 2040 scenarios.  Table~\ref{tab:DAStats} summarizes the associated statistics.   Flexible operation reduced total production costs by 29.3--68.09M\$ or between 1.22--1.76\%.  As illustrated in Figure~\ref{fig:DACost}, Scenarios 2040-2/3/6 have much lower day-ahead production costs due to a high penetration of VREs. In contrast, scenarios 2040-1/4/5 have significantly higher costs as they are forced to commit expensive thermal power plants.  In short, the day-ahead energy market production costs are lower because the flexible mode of operation represents an optimization program that is less constrained than the program associated with the conventional mode of operation.  
\begin{table*}[!h]
\caption{Change in day-ahead energy market production cost statistics (flexible \textit{minus} conventional) for all six 2040 scenarios.}
\begin{footnotesize}
\begin{center}
\begin{tabular}{p{4.2cm}p{1.5cm}p{1.5cm}p{1.5cm}p{1.5cm}p{1.5cm}p{1.5cm}}\toprule
\textbf{$\Delta$ Day-Ahead Costs} & \textbf{2040-1}& \textbf{2040-2}& \textbf{2040-3}& \textbf{2040-4}& \textbf{2040-5}& \textbf{2040-6}\\\toprule
\textbf{Mean} (\$/hr)& 6115.1 \newline(1.22\%) & 5909.4 \newline(1.49\%) & 3345.2 \newline(1.76\%) & 7712.7 \newline(1.41\%) & 7773.1 \newline(1.49\%) & 4388.1 \newline(1.64\%)  \\\midrule
\textbf{STD} (\$/hr) & 4859.0 \newline(2.09\%) & 4355.7 \newline(1.89\%) & 5336.3 \newline(3.89\%) & 5327.3 \newline(2.62\%) & 6160.9 \newline(3.05\%) & 6095.2 \newline(3.02\%) \\\midrule
\textbf{Max} (\$/hr)& -16071.5 \newline(-0.95\%)  & 38820.1 \newline(2.65\%)  & 66093.4 \newline(5.44\%)  & -76701.8 \newline(-4.56\%)  & 15683.0 \newline(0.83\%)  & 476535.0 \newline(23.20\%)  \\\midrule
\textbf{Min} (\$/hr)& 19290.1 \newline(18.95\%) & -2738.0 \newline(-3.14\%) & 15922.7 \newline(19.18\%) & -706.4 \newline(-0.45\%) & -419.0 \newline(-0.36\%) & -10860.0 \newline(-12.17\%)  \\\midrule
\textbf{Total (million \$)} & 53.57 & 51.77 & 29.30 & 67.56 & 68.09 & 38.44  \\\bottomrule
\textbf{\% Reduction} &  1.22 &  1.49 &  1.76 &  1.41 &  1.49 &  1.64 \\\bottomrule
\end{tabular}
\end{center}
\end{footnotesize}

\label{tab:DAStats}
\end{table*}
\newpage
\subsubsection{Real-Time Energy Market Production Costs}
\begin{figure}[!h]
\centering
\includegraphics[width=6.2in]{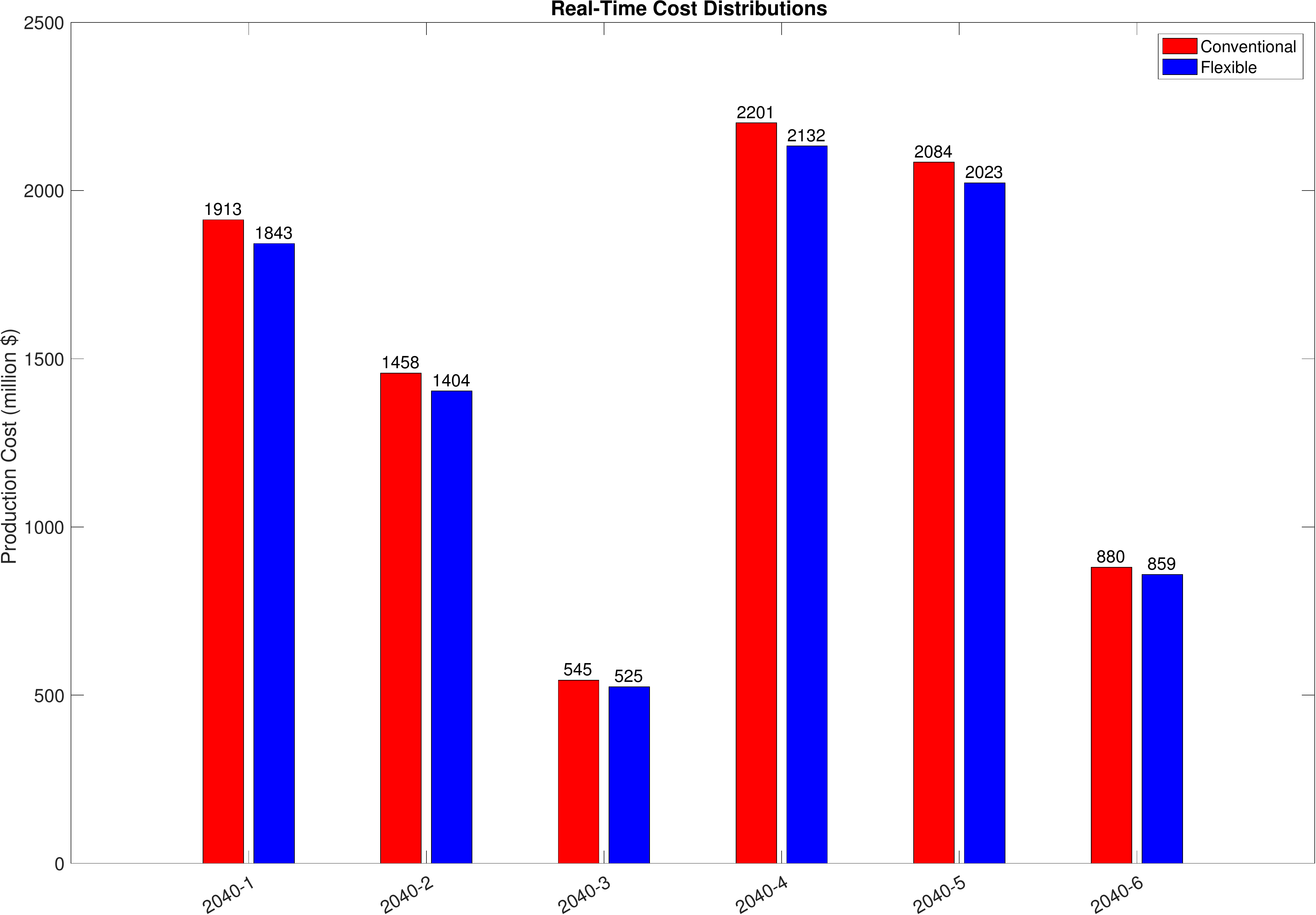}
\caption{A comparison of the real-time production costs for flexible and conventional operation.}
\label{fig:SCEDCost}
\end{figure}

Figure~\ref{fig:SCEDCost} illustrates the total real-time energy market production cost for all six scenarios. Similar to the day-ahead energy market, Scenarios 2040-1/4/5 have significantly higher production costs as they are forced to dispatch more expensive thermal power plants. Meanwhile, Scenarios 2040-2/3/6 have lower real-time energy market production costs due to a greater utilization of renewable energy. As detailed in Table~\ref{tab:SCEDStats}, flexible operation reduces the average real-time market production costs by 2.46\%--3.70\%  (or 19.58-70.83M\$) across all six scenarios. 
\begin{table*}[!h]
\caption{A summary of the real-time production cost statistics (flexible \textit{minus} conventional).}
\begin{footnotesize}
\begin{center}
\begin{tabular}{p{3.5cm}p{1.5cm}p{1.5cm}p{1.5cm}p{1.5cm}p{1.5cm}p{1.5cm}}\toprule
\textbf{$\Delta$ Real-Time Cost} & \textbf{2040-1}& \textbf{2040-2}& \textbf{2040-3}& \textbf{2040-4}& \textbf{2040-5}& \textbf{2040-6}\\\toprule
\textbf{Mean} (\$/min)& 1347.5 \newline(3.70\%) & 1013.5 \newline(3.65\%) & 372.5 \newline(3.59\%) & 1304.9 \newline(3.12\%) & 1173.1 \newline(2.96\%) & 412.5 \newline(2.46\%) \\\midrule
\textbf{STD} (\$/min) & 493.5 \newline(2.31\%) & 533.2 \newline(2.62\%) & 553.8 \newline(5.21\%) & 497.8 \newline(2.58\%) & 545.8 \newline(2.90\%) & 536.9 \newline(3.30\%) \\\midrule
\textbf{Max} (\$/min)& 895.8 \newline(0.58\%) & 3976.9 \newline(2.69\%) & 385.2 \newline(0.36\%) & 3163.4 \newline(2.02\%) & -5845.8 \newline(-3.41\%) & 40662.3 \newline(23.52\%) \\\midrule
\textbf{ Min} (\$/min)& 88.4 \newline(2.76\%) & 75.5 \newline(3.45\%) & -0.0 \newline(-0.00\%) & 65.3 \newline(0.98\%) & -0.0 \newline(-0.00\%) & 157.3 \newline(3.78\%) \\\bottomrule
\textbf{Total (million \$)} & 70.83 & 53.27 & 19.58 & 68.58 & 61.66 & 21.7  \\\bottomrule
\textbf{\% Reduction} &  3.70 &  3.65 &  3.59 &  3.12 &  2.96 &  2.46 \\\bottomrule
\end{tabular}
\end{center}
\end{footnotesize}

\label{tab:SCEDStats}
\end{table*}

\section{Conclusion}\label{sec:conc}
This work has used a novel enterprise control assessment methodology to study the energy-water nexus for the ISO New England System.  Six scenarios were studied representing plausible electric power capacity mixes in 2040.  The study specifically sought to understand the impact of flexible coordinated operation of energy-water resources on the holistic behavior of these six scenarios.  In short, the flexible operation energy-water resources demonstrated truly ``sustainable synergies'' with respect to balancing, environmental, and economic performance.   Table \ref{tab:scorecard} summarizes the most important results of the study in a balanced sustainability scorecard and highlights the synergistic improvements caused by flexible coordinated operation of the energy-water nexus.  These results show that as VRE resources become an ever-important part of the electric power system landscape, so too must the electric power system evolve to engage energy-water resources as control levers.  In some cases, such resources -- like hydro-power plants -- are mainstays of traditional operation.  In other cases, particularly water utility electric loads, these resources will have to evolve their operation to become true electric power grid participants.  

\begin{table*}[!h]
\caption{Balanced Sustainability Scorecard:  The range of \textbf{\emph{improvements}} caused by coordinated flexible operation of the energy-water nexus.}
\begin{footnotesize}
\begin{center}
\begin{tabular}{p{7cm}p{3cm}}\toprule
\textbf{Balancing Performance}\\\toprule
Average Load Following Reserves& 1.24--12.66\% \\\midrule
Average Ramping Reserves& 5.28--18.35\% \\\midrule
Percent Time Curtailed & 2.67--10.90\% \\\midrule
Percent Time Exhausted Regulation Reserves & 0\% \\\midrule
Std. Dev. of Imbalances & 3.874--6.484\% \\\midrule
\textbf{Environmental Performance} \\\toprule
Total Water Withdrawals & 0.65--25.58\% \\\midrule
Total Water Consumption & 1.03--5.30\% \\\midrule
Total CO$_2$ Emissions & 2.10--3.46\% \\\midrule
\textbf{Economic Performance}\\\toprule
Total Day-Ahead Energy Market Production Cost & 29.30--68.09M\$ \\\midrule
Total Real-Time Energy Market Production Cost & 19.58--70.83M\$ \\\bottomrule
\end{tabular}
\end{center}
\end{footnotesize}
\label{tab:scorecard}
\end{table*}

\bibliographystyle{IEEEtran}
\bibliography{ewn,LIINESLibrary,LIINESPublications,ISONE}

\begin{thebibliography}{10}
\providecommand{\url}[1]{#1}
\csname url@rmstyle\endcsname
\providecommand{\newblock}{\relax}
\providecommand{\bibinfo}[2]{#2}
\providecommand\BIBentrySTDinterwordspacing{\spaceskip=0pt\relax}
\providecommand\BIBentryALTinterwordstretchfactor{4}
\providecommand\BIBentryALTinterwordspacing{\spaceskip=\fontdimen2\font plus
\BIBentryALTinterwordstretchfactor\fontdimen3\font minus
  \fontdimen4\font\relax}
\providecommand\BIBforeignlanguage[2]{{%
\expandafter\ifx\csname l@#1\endcsname\relax
\typeout{** WARNING: IEEEtran.bst: No hyphenation pattern has been}%
\typeout{** loaded for the language `#1'. Using the pattern for}%
\typeout{** the default language instead.}%
\else
\language=\csname l@#1\endcsname
\fi
#2}}

\bibitem{Farid:2016:00}
\BIBentryALTinterwordspacing
A.~M. Farid, ``The liines commitment to open-information,'' The Labortory for
  Intelligent Integrated Networks of Engineering Systems, Tech. Rep., 2016.
  [Online]. Available:
  \url{http://engineering.dartmouth.edu/liines/wpblog/2016/02/07/the-liines-commitment-to-open-information/}
\BIBentrySTDinterwordspacing

\bibitem{EPRI:2016:00}
{EPRI}, ``Electric power system flexibility: Challenges and opportunities,''
  Electric Power Research Institute, Tech. Rep. 3002007374, February 2016.

\bibitem{ISO-NE:2017:02}
\BIBentryALTinterwordspacing
ISO-NE. (2019) {Resource Mix-ISO New England}. [Online]. Available:
  \url{https://www.iso-ne.com/about/key-stats/resource-mix}
\BIBentrySTDinterwordspacing

\bibitem{Sanders:2014:00}
K.~T. Sanders, M.~F. Blackhurst, C.~W. King, and M.~E. Webber, ``{The Impact of
  Water Use Fees on Dispatching and Water Requirements for Water-Cooled Power
  Plants in Texas},'' \emph{{Environ. Sci. Technol.}}, vol.~48, no.~12, p.
  140602120931006, Jun 2014.

\bibitem{Armstrong:2018:00}
N.~R. Armstrong, R.~C. Shallcross, K.~Ogden, S.~Snyder, A.~Achilli, and E.~L.
  Armstrong, ``Challenges and opportunities at the nexus of energy, water, and
  food: A perspective from the southwest united states,'' \emph{MRS Energy \&
  Sustainability}, vol.~5, 2018.

\bibitem{Muzhikyan:2019:SPG-JR04}
A.~Muzhikyan, S.~Muhanji, G.~Moynihan, D.~Thompson, Z.~Berzolla, and A.~M.
  Farid, ``{The 2017 ISO New England System Operational Analysis and Renewable
  Energy Integration Study},'' \emph{{Energy Reports}}, vol.~5, pp. 747--792,
  {July} 2019.

\bibitem{Rogers:2013:00}
J.~Rogers, K.~Averyt, S.~Clemmer, M.~Davis, F.~Flores-Lopez, D.~Kenney,
  J.~Macknick, N.~Madden, J.~Meldrum, S.~Sattler, and E.~Spanger-Siegfried,
  ``{Water-Smart Power: Strengthening the U.S. Electricity System in a Warming
  World},'' Union for Concerned Scientists, Cambridge, MA, Tech. Rep., 2013.

\bibitem{Kanyerere:2018:00}
T.~Kanyerere, S.~Tramberend, A.~D. Levine, P.~Mokoena, P.~Mensah, W.~Chingombe,
  J.~Goldin, S.~Fatima, and M.~Prakash, ``Water futures and solutions: Options
  to enhance water security in sub-saharan africa,'' in \emph{Systems Analysis
  Approach for Complex Global Challenges}.\hskip 1em plus 0.5em minus
  0.4em\relax Springer, 2018, pp. 93--111.

\bibitem{Averyt:2013:00}
K.~Averyt, J.~Macknick, J.~Rogers, N.~Madden, J.~Fisher, J.~Meldrum, and
  R.~Newmark, ``Water use for electricity in the united states: an analysis of
  reported and calculated water use information for 2008,'' \emph{Environmental
  Research Letters}, vol.~8, no.~1, p. 015001, Jan 2013.

\bibitem{Muhanji:2019:SPG-JR05}
\BIBentryALTinterwordspacing
S.~O. Muhanji and A.~M. Farid, ``{An Enterprise Control Methodology for the
  Techno-Economic Assessment of the Energy Water Nexus},'' \emph{(under
  revision)}, vol.~1, no.~1, p.~31, [SPG-JR05] 2019. [Online]. Available:
  \url{https://arxiv.org/abs/1908.10469}
\BIBentrySTDinterwordspacing

\bibitem{Al-Nory:2014:00}
M.~Al-Nory and M.~El-Beltagy, ``An energy management approach for renewable
  energy integration with power generation and water desalination,''
  \emph{Renewable Energy}, vol.~72, pp. 377--385, 2014.

\bibitem{Lubega:2016:00}
W.~N. Lubega and A.~M. Farid, ``A reference system architecture for the
  energy--water nexus,'' \emph{{IEEE Systems Journal}}, vol.~10, no.~1, pp.
  106--116, 2016.

\bibitem{Meldrum:2013:00}
J.~Meldrum, S.~Nettles-Anderson, G.~Heath, and J.~Macknick, ``Life cycle water
  use for electricity generation: a review and harmonization of literature
  estimates,'' \emph{Environmental Research Letters}, vol.~8, no.~1, p. 015031,
  Mar 2013.

\bibitem{Macknick:2012:00}
J.~Macknick, R.~Newmark, G.~Heath, and K.~C. Hallett, ``Operational water
  consumption and withdrawal factors for electricity generating technologies: a
  review of existing literature,'' \emph{Environmental Research Letters},
  vol.~7, no.~4, p. 045802, Dec 2012.

\bibitem{Averyt:2011:00}
K.~Averyt, J.~Fisher, A.~Huber-Lee, A.~Lewis, J.~Macknick, N.~Madden,
  J.~Rogers, and S.~Tellinghuisen, ``Freshwater use by us power plants:
  Electricity's thirst for a precious resource,'' Union of Concerned
  Scientists, Cambridge, MA, USA, Tech. Rep., 2011.

\bibitem{Macknick:2012:01}
J.~Macknick, S.~Sattler, K.~Averyt, S.~Clemmer, and J.~Rogers, ``The water
  implications of generating electricity: water use across the united states
  based on different electricity pathways through 2050,'' \emph{Environmental
  Research Letters}, vol.~7, no.~4, p. 045803, Dec 2012.

\bibitem{Bagloee:2018:00}
S.~A. Bagloee, M.~Asadi, and M.~Patriksson, ``Minimization of water pumps'
  electricity usage: a hybrid approach of regression models with
  optimization,'' \emph{Expert Systems with Applications}, 2018.

\bibitem{Bagirov:2013:00}
A.~M. Bagirov, A.~Barton, H.~Mala-Jetmarova, A.~Al~Nuaimat, S.~Ahmed,
  N.~Sultanova, and J.~Yearwood, ``An algorithm for minimization of pumping
  costs in water distribution systems using a novel approach to pump
  scheduling,'' \emph{Mathematical and Computer Modelling}, vol.~57, no. 3-4,
  pp. 873--886, 2013.

\bibitem{Ulanicki:2007:00}
B.~Ulanicki, J.~Kahler, and H.~See, ``Dynamic optimization approach for solving
  an optimal scheduling problem in water distribution systems,'' \emph{Journal
  of Water Resources Planning and Management}, vol. 133, no.~1, pp. 23--32,
  2007.

\bibitem{Lopez-Ibanez:2008:00}
M.~L{\'o}pez-Ib{\'a}{\~n}ez, T.~D. Prasad, and B.~Paechter, ``Ant colony
  optimization for optimal control of pumps in water distribution networks,''
  \emph{Journal of Water Resources Planning and Management}, vol. 134, no.~4,
  pp. 337--346, 2008.

\bibitem{Ghelichi:2018:00}
Z.~Ghelichi, J.~Tajik, and M.~S. Pishvaee, ``A novel robust optimization
  approach for an integrated municipal water distribution system design under
  uncertainty: A case study of mashhad,'' \emph{Computers \& Chemical
  Engineering}, vol. 110, pp. 13--34, 2018.

\bibitem{Diaz:2017:00}
C.~Diaz, F.~Ruiz, and D.~Patino, ``Modeling and control of water booster
  pressure systems as flexible loads for demand response,'' \emph{Applied
  Energy}, vol. 204, pp. 106--116, 2017.

\bibitem{Takahashi:2017:00}
S.~Takahashi, H.~Koibuchi, and S.~Adachi, ``Water supply operation and
  scheduling system with electric power demand response function,''
  \emph{Procedia Engineering}, vol. 186, pp. 327--332, 2017.

\bibitem{Menke:2017:00}
R.~Menke, E.~Abraham, P.~Parpas, and I.~Stoianov, ``Extending the envelope of
  demand response provision though variable speed pumps,'' \emph{Procedia
  Engineering}, vol. 186, pp. 584--591, 2017.

\bibitem{Menke:2016:01}
------, ``Demonstrating demand response from water distribution system through
  pump scheduling,'' \emph{Applied Energy}, vol. 170, pp. 377--387, 2016.

\bibitem{Santhosh:2013:EWN-C16}
\BIBentryALTinterwordspacing
A.~Santhosh, A.~M. Farid, and K.~Youcef-Toumi, ``{Optimal Network Flow for the
  Supply Side of the Energy-Water Nexus},'' in \emph{2013 IEEE International
  Workshop on Intelligent Energy Systems}, Vienna, Austria, 2013, pp. 1--6.
  [Online]. Available:
  \url{http://dx.doi.org.libproxy.mit.edu/10.1109/IWIES.2013.6698578}
\BIBentrySTDinterwordspacing

\bibitem{Santhosh:2012:EWN-C09}
\BIBentryALTinterwordspacing
A.~Santhosh, A.~M. Farid, A.~Adegbege, and K.~Youcef-Toumi, ``{Simultaneous
  Co-optimization for the Economic Dispatch of Power and Water Networks},'' in
  \emph{The 9th IET International Conference on Advances in Power System
  Control, Operation and Management}, Hong Kong, China, 2012, pp. 1--6.
  [Online]. Available: \url{http://dx.doi.org/10.1049/cp.2012.2148}
\BIBentrySTDinterwordspacing

\bibitem{Hickman:2017:EWN-J32}
\BIBentryALTinterwordspacing
W.~Hickman, A.~Muzhikyan, and A.~M. Farid, ``{The Synergistic Role of Renewable
  Energy Integration into the Unit Commitment of the Energy Water Nexus},''
  \emph{Renewable Energy}, vol. 108, no.~1, pp. 220--229, 2017. [Online].
  Available: \url{https://dx.doi.org/10.1016/j.renene.2017.02.063}
\BIBentrySTDinterwordspacing

\bibitem{Ela:2009:00}
E.~Ela, M.~Milligan, B.~Parsons, D.~Lew, and D.~Corbus, ``The evolution of wind
  power integration studies: past, present, and future,'' in \emph{Power \&
  Energy Society General Meeting, 2009. PES'09. IEEE}.\hskip 1em plus 0.5em
  minus 0.4em\relax IEEE, 2009, pp. 1--8.

\bibitem{Brouwer:2014:00}
A.~S. Brouwer, M.~van~den Broek, A.~Seebregts, and A.~Faaij, ``{Impacts of
  large-scale Intermittent Renewable Energy Sources on electricity systems ,
  and how these can be modeled},'' \emph{Renewable and Sustainable Energy
  Reviews}, vol.~33, pp. 443--466, 2014.

\bibitem{Holttinen:2012:01}
H.~Holttinen, M.~O. Malley, J.~Dillon, and D.~Flynn, ``Recommendations for wind
  integration studies -- {IEA} task 25,'' International Energy Agency,
  Helsinki, Tech. Rep., 2012.

\bibitem{Holttinen:2013:00}
H.~Holttinen, A.~Orths, H.~Abilgaard, F.~van Hulle, J.~Kiviluoma, B.~Lange,
  M.~OMalley, D.~Flynn, A.~Keane, J.~Dillon, E.~M. Carlini, J.~O. Tande,
  A.~Estanquiro, E.~G. Lazaro, L.~Soder, M.~Milligan, C.~Smith, and C.~Clark,
  ``Iea wind export group report on recommended practices wind integration
  studies,'' International Energy Agency, Paris, France, Tech. Rep., 2013.

\bibitem{GE-Energy:2010:01}
GE-Energy, ``New england wind integration study,'' GE Energy and ISO New
  England, Schenectady, New York, Tech. Rep. May, 2010.

\bibitem{Shlatz:2011:00}
E.~Shlatz, L.~Frantzis, T.~McClive, G.~Karlson, D.~Acharya, S.~Lu, P.~Etingov,
  R.~Diao, J.~Ma, N.~Samaan, V.~Chadliev, M.~Smart, R.~Salgo, R.~Sorensen,
  B.~Allen, B.~Idelchik, A.~Ellis, J.~Stein, C.~Hanson, Y.~V. Makarov, X.~Guo,
  R.~P. Hafen, C.~Jin, and H.~Kirkham, ``Large-scale pv integration study,''
  Navigant Consulting, Las Vegas, NV, USA, Tech. Rep., 2011.

\bibitem{Piwko:2005:00}
R.~Piwko, X.~Bai, K.~Clark, G.~Jordan, N.~Miller, and J.~Zimberlin, ``{The
  Effects of Integrating Wind Power on Transmission System Planning,
  Reliability and Operations},'' GE Energy, Schnectady, New York, Tech. Rep.,
  2005.

\bibitem{Farid:2014:SPG-J26}
\BIBentryALTinterwordspacing
A.~M. Farid, B.~Jiang, A.~Muzhikyan, and K.~Youcef-Toumi, ``{The Need for
  Holistic Enterprise Control Assessment Methods for the Future Electricity
  Grid},'' \emph{Renewable \& Sustainable Energy Reviews}, vol.~56, no.~1, pp.
  669--685, 2015. [Online]. Available:
  \url{http://dx.doi.org/10.1016/j.rser.2015.11.007}
\BIBentrySTDinterwordspacing

\bibitem{Holttinen:2008:01}
H.~Holttinen, M.~Milligan, B.~Kirby, T.~Acker, V.~Neimane, and T.~Molinski,
  ``Using standard deviation as a measure of increased operational reserve
  requirement for wind power by wind engineering using standard deviation as a
  measure of increased operational reserve requirement for wind power,''
  vol.~44, no.~0, 2008.

\bibitem{Robitaille:2012:00}
A.~Robitaille, I.~Kamwa, A.~H. Oussedik, M.~de~Montigny, N.~Menemenlis,
  M.~Huneault, A.~Forcione, R.~Mailhot, J.~Bourret, and L.~Bernier,
  ``Preliminary impacts of wind power integration in the hydro-quebec system,''
  \emph{Wind Engineering}, vol.~36, no.~1, pp. 35--52, Feb 2012.

\bibitem{Ummels:2007:00}
\BIBentryALTinterwordspacing
B.~C. Ummels, M.~Gibescu, E.~Pelgrum, W.~L. Kling, and A.~J. Brand, ``{Impacts
  of Wind Power on Thermal Generation Unit Commitment and Dispatch},''
  \emph{IIEEE Transactions on Energy Conversion}, vol.~22, no.~1, pp. 44--51,
  2007. [Online]. Available: \url{http://dx.doi.org/10.1109/TEC.2006.889616
  http://ieeexplore.ieee.org/lpdocs/epic03/wrapper.htm?arnumber=4106021}
\BIBentrySTDinterwordspacing

\bibitem{Curtright:2008:01}
A.~E. Curtright and J.~Apt, ``The character of power output from utility-scale
  photovoltaic systems,'' \emph{Progress in Photovoltaics: Research and
  Applications}, vol.~16, no.~3, pp. 241--247, 2008.

\bibitem{Apt:2007:00}
J.~Apt and A.~Curtright, ``The spectrum of power from utility-scale wind farms
  and solar photovoltaic arrays,'' Carnegie Mellon Electricity Industry Center
  Working Paper, Pittsburgh, PA, United states, Tech. Rep., 2007.

\bibitem{Milano:2010:17}
\BIBentryALTinterwordspacing
F.~Milano, \emph{{Power system modelling and scripting}}, 1st~ed.\hskip 1em
  plus 0.5em minus 0.4em\relax New York: Springer, 2010. [Online]. Available:
  \url{http://www.uclm.es/area/gsee/web/Federico/psat.htm}
\BIBentrySTDinterwordspacing

\bibitem{Muzhikyan:2017:SPG-J36}
\BIBentryALTinterwordspacing
A.~Muzhikyan, T.~Mezher, and A.~M. Farid, ``{Power System Enterprise Control
  with Inertial Response Procurement},'' \emph{IEEE Transactions on Power
  Systems (in press)}, vol.~33, no.~4, pp. 3735 -- 3744, 2018. [Online].
  Available: \url{http://dx.doi.org/10.1109/TPWRS.2017.2782085}
\BIBentrySTDinterwordspacing

\bibitem{Muzhikyan:2015:SPG-J22}
\BIBentryALTinterwordspacing
A.~Muzhikyan, A.~M. Farid, and K.~Youcef-Toumi, ``{Relative Merits of Load
  Following Reserves and Energy Storage Market Integration Towards Power System
  Imbalances},'' \emph{International Journal of Electrical Power and Energy
  Systems}, vol.~74, no.~1, pp. 222--229, 2016. [Online]. Available:
  \url{http://dx.doi.org/10.1016/j.ijepes.2015.07.013}
\BIBentrySTDinterwordspacing

\bibitem{Muzhikyan:2015:SPG-J16}
\BIBentryALTinterwordspacing
------, ``{An Enterprise Control Assessment Method for Variable Energy Resource
  Induced Power System Imbalances Part 2: Results},'' \emph{IEEE Transactions
  on Industrial Electronics}, vol.~62, no.~4, pp. 2459 -- 2467, 2015. [Online].
  Available: \url{http://dx.doi.org/10.1109/TIE.2015.2395380}
\BIBentrySTDinterwordspacing

\bibitem{Muzhikyan:2015:SPG-J15}
\BIBentryALTinterwordspacing
------, ``{An Enterprise Control Assessment Method for Variable Energy Resource
  Induced Power System Imbalances Part 1: Methodology},'' \emph{IEEE
  Transactions on Industrial Electronics}, vol.~62, no.~4, pp. 2448--2458,
  2015. [Online]. Available: \url{http://dx.doi.org/10.1109/TIE.2015.2395391}
\BIBentrySTDinterwordspacing

\bibitem{Muzhikyan:2014:SPG-C43}
\BIBentryALTinterwordspacing
------, ``{A Power Grid Enterprise Control Method for Energy Storage System
  Integration},'' in \emph{IEEE Innovative Smart Grid Technologies Conference
  Europe}, Istanbul, Turkey, 2014, pp. 1--6. [Online]. Available:
  \url{http://dx.doi.org/10.1109/ISGTEurope.2014.7028898}
\BIBentrySTDinterwordspacing

\bibitem{Muzhikyan:2014:SPG-AP07}
------, ``{Enterprise Control Mitigation of Variable Energy Resource Induced
  Power System Imbalances},'' in \emph{9th Carnegie Mellon University
  Electricity Conference}, Pittsburgh, PA, United states, 2014, pp. 1--41.

\bibitem{Abdulla:2015:EWN-C53}
\BIBentryALTinterwordspacing
H.~Abdulla and A.~M. Farid, ``Extending the energy-water nexus reference
  architecture to the sustainable development of agriculture, industry \&
  commerce,'' in \emph{First IEEE International Smart Cities Conference},
  Guadalajara, Mexico, 2015, pp. 1--7. [Online]. Available:
  \url{http://dx.doi.org/10.1109/ISC2.2015.7366166}
\BIBentrySTDinterwordspacing

\bibitem{Rutberg:2011:00}
M.~J. Rutberg, A.~Delgado, H.~J. Herzog, and A.~F. Ghoniem, ``A system-level
  generic model of water use at power plants and its application to regional
  water use estimation,'' in \emph{ASME 2011 International Mechanical
  Engineering Congress and Exposition}.\hskip 1em plus 0.5em minus 0.4em\relax
  American Society of Mechanical Engineers, 2011, pp. 513--523.

\bibitem{Cohen:2019:00}
S.~Cohen, J.~Becker, D.~Bielen, M.~Brown, W.~Cole, K.~Eurek, W.~Frazier,
  B.~Frew, P.~Gagnon, J.~Ho, P.~Jadun, T.~Mai, M.~Mowers, C.~Murphy,
  A.~Reimers, J.~Richards, N.~Ryan, E.~Spyrou, D.~Steinberg, Y.~Sun,
  N.~Vincent, and M.~Zwerling, ``{Regional Energy Deployment System (ReEDS)
  Model Documentation: Version 2018},'' {National Renewable Energy Lab. (NREL),
  Golden, CO (United States)}, Tech. Rep. NREL/TP-6A20- 72023, April 2019.

\bibitem{Short:2011:00}
W.~Short, P.~Sullivan, T.~Mai, M.~Mowers, C.~Uriarte, N.~Blair, D.~Heimiller,
  and A.~Martinez, ``{Regional Energy Deployment System (ReEDS)},'' {National
  Renewable Energy Lab.(NREL), Golden, CO (United States)}, Tech. Rep.
  NREL/TP-6A20-46534, December 2011.

\bibitem{Eurek:2016:00}
K.~Eurek, W.~Cole, D.~Bielen, N.~Blair, S.~Cohen, B.~Frew, J.~Ho, V.~Krishnan,
  T.~Mai, B.~Sigrin, \emph{et~al.}, ``{Regional Energy Deployment System
  (ReEDS) Model Documentation: Version 2016},'' {National Renewable Energy
  Lab.(NREL), Golden, CO (United States)}, Tech. Rep. NREL/TP-6A20-67067,
  November 2016.

\bibitem{Cole:2018:00}
W.~Cole, W.~Frazier, P.~Donohoo-Vallett, T.~Mai, and P.~Das, ``{2018 Standard
  Scenarios Report: A U.S. Electricity Sector Outlook},'' {National Renewable
  Energy Lab (NREL)}, Tech. Rep. NREL/TP-6A20-71913, November 2018.

\bibitem{Gomez-Exposito:2011:02}
A.~G{\'o}mez-Exp{\'o}sito, A.~de~la Villa~Ja{\'e}n, C.~G{\'o}mez-Quiles,
  P.~Rousseaux, and T.~Van~Cutsem, ``A taxonomy of multi-area state estimation
  methods,'' \emph{Electric Power Systems Research}, vol.~81, no.~4, pp.
  1060--1069, Apr 2011.

\bibitem{Holttinen:2012:00}
H.~Holttinen, M.~Milligan, E.~Ela, N.~Menemenlis, J.~Dobschinski, B.~Rawn,
  R.~J. Bessa, D.~Flynn, E.~Gomez-Lazaro, and N.~K. Detlefsen, ``Methodologies
  to determine operating reserves due to increased wind power,'' \emph{IEEE
  Trans. Sustain. Energy}, vol.~3, no.~4, pp. 713--723, Oct 2012.

\bibitem{Ela:2010:00}
E.~Ela, B.~Kirby, E.~Lannoye, M.~Milligan, D.~Flynn, B.~Zavadil, and
  M.~O'Malley, ``Evolution of operating reserve determination in wind power
  integration studies,'' in \emph{Power and Energy Society General Meeting,
  2010 IEEE}.\hskip 1em plus 0.5em minus 0.4em\relax IEEE, 2010, pp. 1--8.

\bibitem{Rutberg:2012:00}
M.~J. Rutberg, ``Modeling water use at thermoelectric power plants,'' Ph.D.
  dissertation, Massachussets Institute of Technology, 2012.

\bibitem{Integrated-Pollution-Prevention-and-Control-IPPC:2001:00}
{Integrated Pollution Prevention and Control (IPPC)} and E.~Commission,
  ``Reference document on the application of best available techniques to
  industrial cooling systems december 2001,'' European Commission, Tech. Rep.
  {December}, 2001.

\bibitem{ECOFYS:2014:00}
ECOFYS, ``-- geo-localised inventory of water use in cooling processes,
  assessment of vulnerability and of water use management measures,'' European
  Commission, End Report ENV.D1/SER/2013/0004, {July} 2014.

\bibitem{Eurelectric:1999:00}
Eurelectric, ``Bat for cooling systems, working group "environmental
  protection",'' Eurelectric, Tech. Rep., 1999.

\bibitem{Tsou:2013:00}
J.~L. Tsou, J.~Maulbetsch, J.~Shi, and EPRI, ``Power plant cooling system
  overview for researchers and technology developers,'' EPRI, Tech. Rep., 2013.

\bibitem{Pan:2018:01}
S.-Y. Pan, S.~W. Snyder, A.~I. Packman, Y.~J. Lin, and P.-C. Chiang, ``Cooling
  water use in thermoelectric power generation and its associated challenges
  for addressing water-energy nexus,'' \emph{Water-Energy Nexus}, vol.~1,
  no.~1, pp. 26--41, 2018.

\bibitem{EIA923:2019:00}
\BIBentryALTinterwordspacing
E.~I.~A. (EIA). Form eia-923 detailed data with previous form data
  (eia-906/920). [Online]. Available:
  \url{https://www.eia.gov/electricity/data/eia923/}
\BIBentrySTDinterwordspacing

\bibitem{EIA860:2019:00}
\BIBentryALTinterwordspacing
{Energy Information Agency (EIA)}. Form eia-860 detailed data with previous
  form data (eia-860a/860b). [Online]. Available:
  \url{https://www.eia.gov/electricity/data/eia860/}
\BIBentrySTDinterwordspacing

\bibitem{EIA:2019:00}
\BIBentryALTinterwordspacing
(2019) Form eia-767 historical data files. [Online]. Available:
  \url{https://www.eia.gov/electricity/data/eia767/}
\BIBentrySTDinterwordspacing

\end{thebibliography}

\end{document}